\documentclass[aps, superscriptaddress, twocolumn,10pt]{revtex4-2}

\usepackage{graphicx}    
\usepackage{amsmath,amssymb} 
\usepackage{bm}             
\usepackage{physics}       
\usepackage{hyperref}
\hypersetup{
	colorlinks=true,   
	linkcolor=red,  
	citecolor=blue, 
	urlcolor=blue  
}

\begin{document}
	\title{Scalable Repeater Architecture for Long-Range Quantum Energy Teleportation in Gapped Systems}
	\author{M. Y. Abd-Rabbou}~
\email{m.elmalky@azhar.edu.eg}
\affiliation{School of Physics, University of Chinese Academy of Sciences, Yuquan Road 19A, Beijing, 100049, China.}
\affiliation{Mathematics Department, Faculty of Science, Al-Azhar University, Nasr City 11884, Cairo, Egypt.}

\author{Irfan Siddique}
\email{irfansiddique@ucas.ac.cn}
\affiliation{College of Mathematics and Physics, College of Nuclear Energy Science and Engineering, China Three Gorges University, Yichang 443002, China.}
\affiliation{Center for Astronomy and Space Sciences, China Three Gorges University, Yichang 443002, China.}
\affiliation{School of Nuclear Science and Technology, University of Chinese Academy of Sciences 101408, Beijing, China.}

\author{Saeed Haddadi}
\email{haddadi@ipm.ir}
\affiliation{School of Particles and Accelerators, Institute for Research in Fundamental Sciences (IPM), P.O. Box 19395-5531, Tehran, Iran.}

\author{Cong‑Feng Qiao}
\email{qiaocf@ucas.ac.cn}
\affiliation{School of Physics, University of Chinese Academy of Sciences, Yuquan Road 19A, Beijing, 100049, China.}
\affiliation{International Centre for Theoretical Physics Asia-Pacific, UCAS, Beijing 100190, China.}

	\begin{abstract}
		Quantum Energy Teleportation (QET) constitutes a paradigm-shifting protocol that permits the activation of local vacuum energy through the consumption of pre-existing entanglement and classical communication. Nevertheless, the implementation of QET is severely impeded by the fundamental locality of gapped many-body systems, where the exponential clustering of ground-state correlations restricts energy extraction to microscopic scales. In this work, we address this scalability crisis within the framework of the one-dimensional anisotropic XY model. We initially provide a rigorous characterization of a monolithic measurement-induced strategy, demonstrating that while bulk projective measurements can theoretically induce long-range couplings, the approach is rendered physically untenable by exponentially diverging thermodynamic costs and vanishing success probabilities. To circumvent this impasse, we propose and analyze a hierarchical quantum repeater architecture adapted for energy teleportation. By orchestrating heralded entanglement generation, iterative entanglement purification, and nested entanglement swapping, our protocol effectively counteracts the fidelity degradation inherent in noisy quantum channels. We establish that this architecture fundamentally alters the operational resource scaling from exponential to polynomial. This proves, for the first time, the physical permissibility and computational tractability of activating vacuum energy at arbitrary distances. The significance lies not in net energy gain, but in establishing long-range QET as a viable protocol for remote quantum control and resource distribution.
	\end{abstract}
	
	\maketitle

	\section{Introduction}
	\label{sec:intro}
	
	The realization of distributed quantum networks necessitates not only the reliable transmission of information but also the efficient management of energy resources across macroscopic distances. In this context, Quantum Energy Teleportation (QET) has emerged as a pivotal protocol, enabling the activation of local energy at a remote node by consuming pre-existing entanglement and classical communication~\cite{hotta2009quantum, Hotta2010, Hotta2010a}. Unlike physical transport, which is bound by carrier velocity, QET leverages measurement-based operations to extract work from the vacuum state without violating causality or local conservation laws~\cite{Hotta2014, Ikeda2025, Wang2024}. Recent thermodynamic analyses have rigorously framed this process within the resource theories of quantum steering~\cite{Fan2024}, coherence consumption~\cite{Fan2024a, Ikeda2024a}, and entropic inequalities~\cite{SanchezCordova2024}, establishing a solid theoretical foundation for energy manipulation.
	
\begin{figure}[t]
		\centering
		\includegraphics[width=\linewidth]{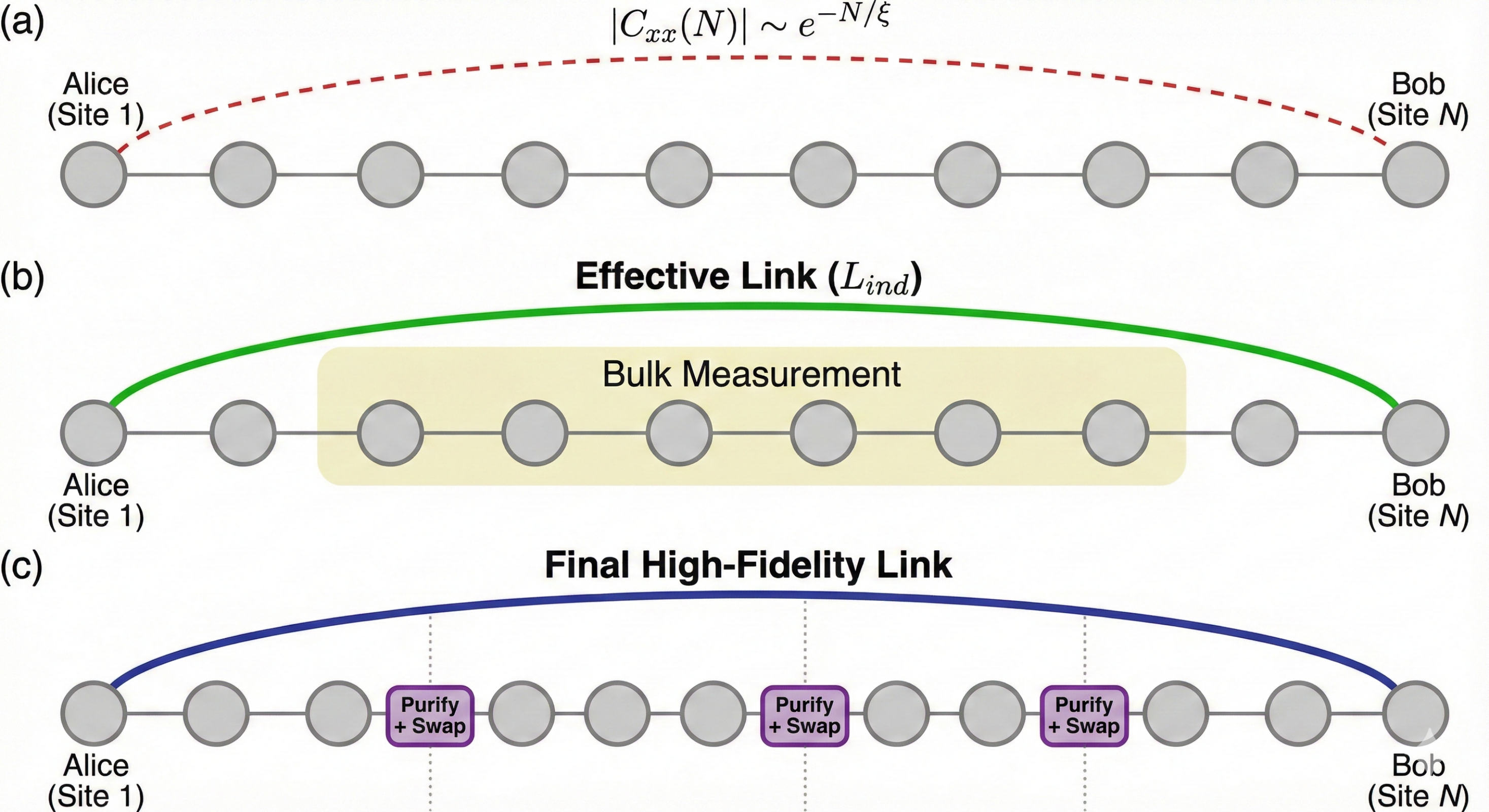} 
		\caption{Scalability in QET. (a) The physical constraint: Extractable energy is strictly bounded by exponentially decaying correlations $C(r)$. (b) The monolithic limitation: Direct measurement strategies incur exponentially diverging energy costs to combat fidelity loss. (c) The repeater solution: A hierarchical architecture utilizing swapping and purification ensures polynomial resource scaling, making long-range energy extraction thermodynamically viable.}
		\label{fig:concept}
	\end{figure}

	However, a critical disparity exists between the theoretical promise of QET and its operational feasibility over long distances. The central motivation for this study arises from the prohibitive thermodynamic cost imposed by the locality of physical interactions. As visualized in Fig.~\ref{fig:concept}(a), ground-state correlations in gapped many-body systems are governed by exponential clustering theorems, leading to a rapid decay of extractable energy as spatial separation increases~\cite{Xie2025, Wu2024, Trevison2015}. Consequently, standard protocols are confined to microscopic regimes~\cite{Frey2013, Yusa2011}, and even enhancements exploiting critical phase transitions fail to provide the robust, distance-independent links required for practical networking~\cite{Ikeda2023, Matsueda2025}.
	
	Attempts to extend the range of QET using single-shot, global measurement strategies—herein referred to as monolithic protocols—are fundamentally flawed. While experimental validations on superconducting hardware \cite{Ikeda2023a} and theoretical models on hyperbolic graphs \cite{Ikeda2024} have demonstrated basic principles, they do not account for the exponential accumulation of errors and energy injection required to bridge macroscopic gaps directly. As depicted in Fig.~\ref{fig:concept}(b), such direct approaches inevitably encounter a point where the energy cost of establishing the link exceeds the energy retrievable, rendering the process thermodynamically futile.
	
	To overcome this intrinsic scalability barrier, we introduce a hierarchical Quantum Repeater (QR) architecture specifically adapted for energy distribution. By synthesizing entanglement swapping and purification techniques originally developed for quantum key distribution~\cite{Fiorini2025, Namiki2016}, we construct a segmented channel that effectively decouples the extraction distance from the correlation length, as shown in Fig.~\ref{fig:concept}(c). This architecture is versatile, supporting implementations in cold-atom ensembles~\cite{Solmeyer2016}, continuous-variable systems~\cite{Wu2022, Dias2017}, and microwave domains~\cite{DiCandia2015}, while employing advanced modulation schemes to maintain high fidelity~\cite{Goncharov2023}. Furthermore, by optimizing the network topology~\cite{Mylavarapu2025} and minimizing entanglement consumption~\cite{Ghosal2025}, our approach ensures that the resource overhead scales polynomially rather than exponentially.
	
	In this paper, we provide a comprehensive blueprint for scalable long-range QET within the anisotropic XY model. Following an analysis of the physical model and the limitations of standard methods (Sec.~\ref{sec:model_and_failure}), we systematically demonstrate that replacing monolithic measurement schemes (Sec.~\ref{sec:monolithic_protocol}) with a repeater-based infrastructure (Sec.~\ref{sec:repeater_protocol}) is the only physically permissible path to sustain non-vanishing energy yields over arbitrary distances. Finally, we describe the complete implementation of this protocol in Sec.~\ref{sec:final_protocol}, thereby establishing a new paradigm for quantum energy networks.

	\section{The Physical Model and Failure of Standard QET}
	\label{sec:model_and_failure}
	
	In this section, we establish the theoretical framework for our analysis. We consider a one-dimensional chain of $N$ qubits (spin-1/2 particles) governed by the anisotropic XY Hamiltonian with open boundary conditions. 
	Assuming a ferromagnetic coupling normalized to unity ($J=1$), the Hamiltonian is given by:
	\begin{equation}
		\hat{H} = -\sum_{j=1}^{N-1} \left( \frac{1+\gamma}{2} \hat{\sigma}_{j}^{x} \hat{\sigma}_{j+1}^{x} + \frac{1-\gamma}{2} \hat{\sigma}_{j}^{y} \hat{\sigma}_{j+1}^{y} \right) - h \sum_{j=1}^{N} \hat{\sigma}_{j}^{z},
		\label{eq:hamiltonian}
	\end{equation}
	where $\hat{\sigma}_{j}^{\alpha}$ (for $\alpha \in \{x,y,z\}$) are the Pauli operators at site $j$, $h$ is the transverse magnetic field strength, and $\gamma \in [0, 1]$ is the anisotropy parameter.
	
	The exact solvability of this model is achieved by mapping the spin operators to spinless fermion operators, $\hat{c}_j$ and $\hat{c}_j^\dagger$, via the non-local Jordan-Wigner transformation. The mapping is defined by the relations:
	\begin{align}
		\hat{\sigma}_{j}^{z} &= 1 - 2\hat{c}_j^\dagger \hat{c}_j
			\label{eq:1jw_transform}
        \end{align}
        and
	\begin{align}	\hat{\sigma}_{j}^{-} &= \left( \prod_{l=1}^{j-1} (-\hat{\sigma}_{l}^{z}) \right) \hat{c}_j,
		\label{eq:jw_transform}
	\end{align}
	where $\hat{\sigma}_{j}^{-} = (\hat{\sigma}_{j}^{x} - i\hat{\sigma}_{j}^{y})/2$ is the spin lowering operator. Following the standard Jordan-Wigner transformation \cite{lieb1961two} defined by Eqs. (\ref{eq:1jw_transform}) and (\ref{eq:jw_transform}), the spin operators are mapped to spinless fermions. By substituting these relations into Eq.(\ref{eq:hamiltonian}) and performing standard algebraic simplifications, the Hamiltonian is diagonalized into a quadratic form of free fermions given by
    \small{
	\begin{equation}
		\hat{H} = -\sum_{j=1}^{N-1} \left[ (\hat{c}_j^\dagger \hat{c}_{j+1} + \text{h.c.}) + \gamma(\hat{c}_j^\dagger \hat{c}_{j+1}^\dagger + \text{h.c.}) \right] + 2h \sum_{j=1}^{N} \hat{c}_j^\dagger \hat{c}_j.
		\label{eq:fermionic_hamiltonian}
	\end{equation}}
	The quadratic nature of this Hamiltonian is pivotal; it implies that the system's ground state is a fermionic Gaussian state. 
	Consequently, all ground-state properties, including the correlation functions essential for QET, are fully determined by the two-point correlation functions, which can be efficiently computed using the covariance matrix formalism. This approach enables efficient numerical calculations and analytical insights.

The primary resource for standard QET protocols is the magnitude of two-point correlations between the sender (Alice, site 1) and receiver (Bob, site $N$). We now rigorously derive the asymptotic behavior of the longitudinal correlation function, $C_{xx}(N) = \langle g|\hat{\sigma}^x_1 \hat{\sigma}^x_N |g\rangle$, where $|g\rangle$ is the ground state.

Using the Jordan-Wigner transformation, the spin operators are expressed as long, non-local strings of Majorana operators. The two-spin operator becomes:
\begin{equation}
	\hat{\sigma}^x_1 \hat{\sigma}^x_N = \hat{A}_1 \left( \prod_{l=1}^{N-1} (i\hat{A}_l\hat{B}_l) \right) \hat{A}_N,
\end{equation}
where $\hat{A}_j = \hat{c}^\dagger_j + \hat{c}_j$ and $\hat{B}_j = i(\hat{c}^\dagger_j - \hat{c}_j)$ are Majorana operators. For a Gaussian ground state $|g\rangle$, the expectation value of such a string can be evaluated using Wick's theorem. This theorem reduces the problem to computing the Pfaffian of the matrix of two-point contractions, which for this specific alternating structure, simplifies to the determinant of an $(N-1)\times(N-1)$ Toeplitz matrix $\mathbf{T}$:
\begin{equation}
	C_{xx}(N) = (-1)^{N-1} \det(\mathbf{T}),
			\label{eq:cxx_determinant}
\end{equation}
where the matrix elements are given by the contractions $T_{jk} = \langle g|\hat{B}_j\hat{A}_{k+1}|g\rangle$.

\begin{figure}[t]
	\centering
	\includegraphics[width=\columnwidth,height=6cm]{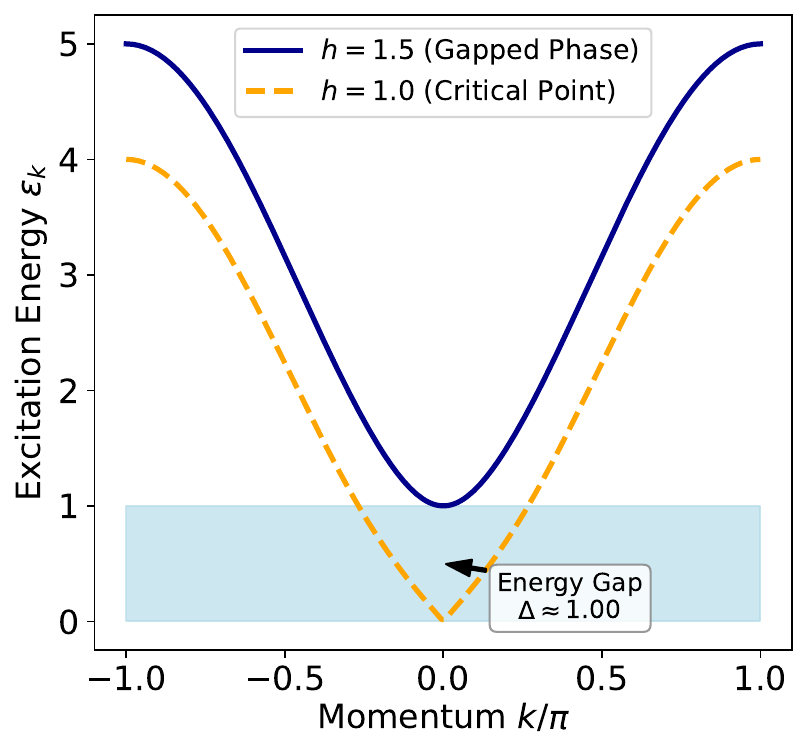}
	\caption{Excitation spectrum and energy gap of the XY model.
		The plot shows the Bogoliubov quasiparticle excitation spectrum $\epsilon_k$ as a function of momentum $k/\pi$. In the paramagnetic phase ($h=1.5 > 1$, solid blue line), a finite energy gap $\Delta = 2(h-1)$ opens at $k=0$. This gap suppresses low-energy excitations and is the direct cause of the exponential decay of ground-state correlations. At the critical point ($h=1.0$, dashed orange line), the gap closes, allowing for the emergence of long-range correlations.}
	\label{fig:energy_gap}
\end{figure}

To understand the physical origin of these correlations, it is necessary to examine the system's excitation spectrum. Diagonalizing the Hamiltonian in Eq. (\ref{eq:fermionic_hamiltonian}) yields the quasiparticle dispersion relation:
	\begin{equation}
		\epsilon_k = 2\sqrt{(h - \cos k)^2 + (\gamma \sin k)^2}.
		\label{eq:spectrum}
	\end{equation}
	As illustrated in Fig. \ref{fig:energy_gap}, for the paramagnetic phase ($h > 1$), this spectrum exhibits a finite energy gap $\Delta = \min_k(\epsilon_k) = 2(h-1)$ at $k=0$. This gap suppresses low-energy fluctuations, which directly dictates the decay behavior of the ground-state correlations.

In the thermodynamic limit ($N \to \infty$) and for the gapped paramagnetic phase ($h > 1$), the asymptotic behavior of the determinant is governed by the Szeg\H{o} limit theorem \cite{PhysRevA.4.2331,bottcher2006analysis}. This theorem dictates that the correlation magnitude decays exponentially with distance $N$:
\begin{equation}
	|C_{xx}(N)| \sim e^{-N/\xi},
			\label{eq:exponential_decay}
\end{equation}
where $\xi$ is the correlation length. By analyzing the analytic properties of the Toeplitz symbol associated with the spectrum in Eq. (\ref{eq:spectrum}), specifically the poles in the complex plane, the correlation length is derived as:
\begin{equation}
	\xi^{-1} = \ln \left( \frac{h + \sqrt{h^2 + \gamma^2 - 1}}{1 + \gamma} \right).
			\label{eq:correlation_length}
\end{equation}
This exponential suppression of correlations is the fundamental barrier to long-range QET. The maximum extractable energy, $E_{\text{std}}$, is bounded by the square of the correlation magnitude, leading to a severe double-exponential decay:
\begin{equation}
	E_{\text{std}}(N) \propto |C_{xx}(N)|^2 \sim e^{-2N/\xi}.
		\label{eq:energy_decay}
\end{equation}
The physical origin of this exponential suppression is the existence of a finite energy gap in the system's excitation spectrum for the paramagnetic phase ($h > 1$), as defined by Eq. (\ref{eq:spectrum}). This gap prevents low-energy, long-wavelength fluctuations, leading to short-range correlations. The rapid decay of usable entanglement renders standard QET protocols ineffective over all but the shortest distances.

	\section{Protocol I: The Monolithic Measurement-Induced Protocol}
	\label{sec:monolithic_protocol}
	
	To circumvent the exponential decay of correlations established in Sec.~\ref{sec:model_and_failure}, we first consider a monolithic protocol. This strategy leverages non-local effects of quantum measurement to create a robust, long-range entangled channel between two distant parties, Alice (site 1) and Bob (site $N$). The protocol proceeds as follows:
	
	\begin{enumerate}
		\item  The system is prepared in the ground state $|g\rangle$ of the Hamiltonian~\eqref{eq:hamiltonian}.
		\item A third party, Charlie, performs a simultaneous, projective measurement on all intermediate bulk qubits, $B = \{2, 3, \dots, N-1\}$.
		\item  The protocol succeeds only if Charlie obtains a specific string of measurement outcomes. Upon success, the measurement collapses the state of the endpoint qubits (Alice and Bob, subsystem $S=\{1,N\}$) into a new, entangled two-qubit state.
	\end{enumerate}

To model the measurement-induced protocol, we employ the covariance matrix formalism for fermionic Gaussian states. The system is fully characterized by the $2N \times 2N$ covariance matrix $\boldsymbol{\Gamma}$ with elements $\Gamma_{lm} = \frac{i}{2} \langle [ \hat{w}_l, \hat{w}_m ] \rangle$, where $\hat{\mathbf{w}} = (\hat{A}_1, \hat{B}_1, \dots, \hat{A}_N, \hat{B}_N)^T$ are Majorana operators.
	Partitioning the system into the endpoints ($S=\{1, N\}$) and the bulk ($B=\{2, \dots, N-1\}$), the matrix is decomposed as:
	\begin{equation}
		\boldsymbol{\Gamma} = \begin{pmatrix} \boldsymbol{\Gamma}_S & \boldsymbol{\Gamma}_{SB} \\ \boldsymbol{\Gamma}_{BS} & \boldsymbol{\Gamma}_B \end{pmatrix}.
			\label{eq:partitioned_cov_matrix}
	\end{equation}
	A projective measurement on the bulk $B$ updates the covariance matrix of the subsystem $S$ according to the Schur complement formula for Gaussian states \cite{RevModPhys.82.277,RevModPhys.84.621}
	\begin{equation}
		\tilde{\boldsymbol{\Gamma}}_S = \boldsymbol{\Gamma}_S - \boldsymbol{\Gamma}_{SB} (\boldsymbol{\Gamma}_B)^{-1} \boldsymbol{\Gamma}_{BS}.
			\label{eq:schur_complement}
	\end{equation}
	This update rule reveals the mechanism for long-range entanglement: although the direct correlations $\boldsymbol{\Gamma}_S$ and cross-correlations $\boldsymbol{\Gamma}_{SB}$ decay exponentially, the inverse bulk matrix $(\boldsymbol{\Gamma}_B)^{-1}$ acts as a long-range propagator. For gapped local Hamiltonians, the inverse of the exponentially decaying banded matrix $\boldsymbol{\Gamma}_B$ exhibits long-range features that compensate for the decay \cite{PhysRevE.85.061126}. Consequently, in the thermodynamic limit, the induced correlation magnitude saturates:
\begin{equation}
	\lim_{N\to\infty} || \mathbf{\Gamma}_{\text{ind}} || = \text{Constant} > 0.
	\label{eq:constant_correlation}
\end{equation}
This proves that the monolithic measurement protocol successfully generates a distance-independent entanglement channel, overcoming the fundamental limitation of exponential decay identified in Sec.~\ref{sec:model_and_failure}. The properties of this induced channel will be analyzed next.

Having established the distance-independent channel, we analyze the sign of the induced correlation $\Gamma_{\text{ind}}$. Using a bond parity argument, the effective link acts as a chain of $N-1$ ferromagnetic bonds. The sign of the induced coupling $\zeta$ is determined by the parity of the total bonds:
\begin{equation}
	\text{sign}(\zeta) \propto (-1)^{N_{\text{bonds}}} = (-1)^{N-1}.
	\label{eq:parity_effect_sign}
\end{equation}
This implies a deterministic zigzag behavior dependent on the chain length $N$: for odd $N$, the coupling is ferromagnetic ($\zeta > 0$); for even $N$, it is antiferromagnetic ($\zeta < 0$).
	To rectify the sign for even $N$, the receiver (Bob) applies a local Pauli-$\hat{\sigma}^z$ correction conditioned on the parity:
\begin{equation}
	\hat{U}_{\text{Bob}} = \begin{cases} \hat{I}_B & \text{if } N \text{ is odd}, \\ \hat{\sigma}^z_B & \text{if } N \text{ is even}. \end{cases}
	\label{eq:bob_correction}
\end{equation}
This unitary operation flips the basis states, effectively reversing the sign of the measured correlations. This simple, deterministic correction ensures that the channel is always ferromagnetic-like, enabling standard energy extraction protocols regardless of distance.

Despite overcoming the correlation decay, the physical viability of this protocol is severely limited by its resource scaling. The success hinges on obtaining a specific outcome string $\mathbf{m}$ from measuring $N-2$ bulk qubits. Since the paramagnetic ground state is highly polarized along the $z$-axis ($|g\rangle \approx |00\dots0\rangle$), a transverse measurement in the $\sigma^y$ basis yields outcomes $\pm 1$ with equal probability $1/2$. The success probability for the entire chain is the product of individual probabilities:
\begin{equation}
	P_{\text{succ}} \approx \prod_{j=2}^{N-1} P(m_j) \approx \left(\frac{1}{2}\right)^{N-2} = 2^{-(N-2)}.
	\label{eq:success_prob_derived}
\end{equation}
This exponentially small probability necessitates an average of $\langle N_{\text{runs}} \rangle = 1/P_{\text{succ}} \sim 2^{N-2}$ experimental runs to achieve a single success.

\begin{figure}[t]
	\centering
\includegraphics[width=\columnwidth,height=5cm]{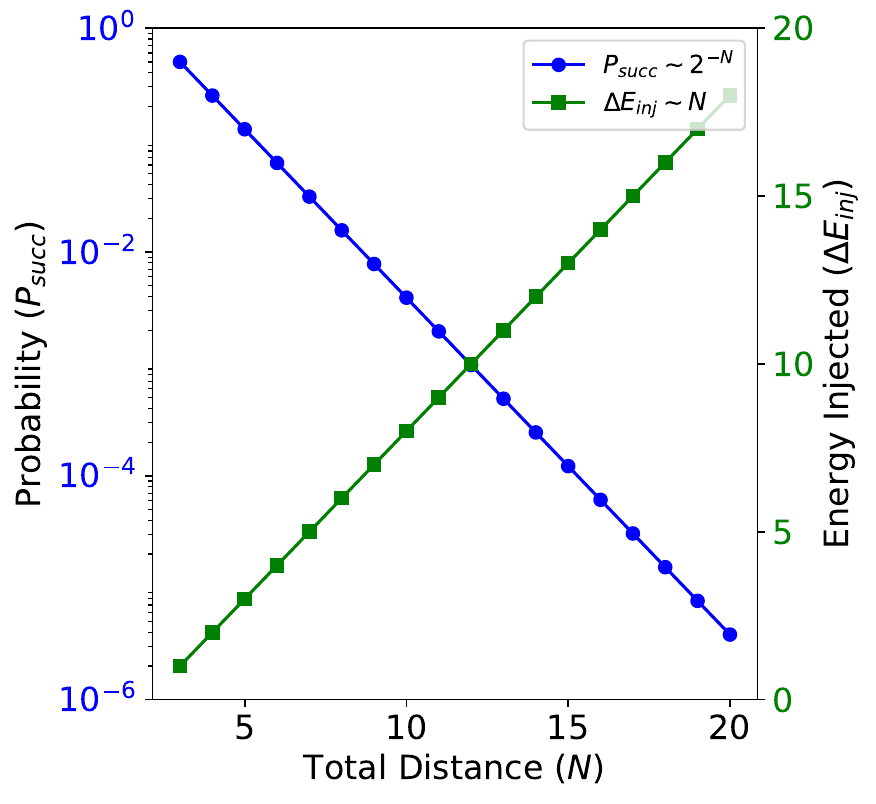}\put(-185,130){$ (a) $}\\
\includegraphics[width=0.88\columnwidth,height=5cm]{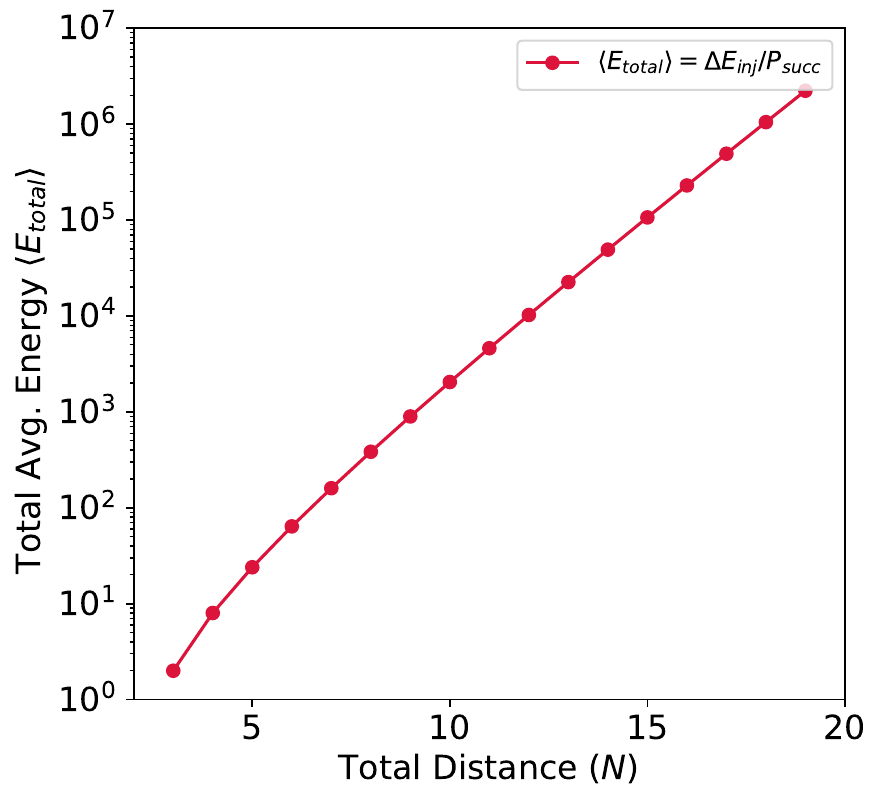}\put(-180,130){$ (b) $}
\caption{Scaling analysis of the monolithic protocol.
	(a)~The two components of the protocol's cost. The success probability, $P_{\text{succ}}$ (blue circles, log scale), decays exponentially with chain length $N$, while the energy injected per attempt, $\Delta E_{\text{inj}}$ (green squares, linear scale), grows linearly. (b)~The resulting total average energy cost, $\langle E_{\text{total}} \rangle$. This cost, which combines the two factors from plot (a), exhibits a prohibitive exponential growth ($\sim N \cdot 2^N$), creating an insurmountable cost wall.}
\label{fig:monolithic_cost_analysis}
\end{figure}

Each run involves a non-unitary projective measurement that injects energy into the system. The initial extensive energy of the ground state is dominated by the alignment with the external field:
\begin{equation}
	E_{\text{pre}} = \langle g|\hat{H}|g\rangle \approx \langle g| -h \sum \hat{\sigma}^z_j |g\rangle \approx -Nh.
\end{equation}
The measurement projects the bulk qubits into eigenstates of $\hat{\sigma}^y$. With respect to the transverse field term $-h\hat{\sigma}^z$, the expectation value of these projected states is zero. Therefore, the post-measurement energy is determined solely by the unmeasured endpoints:
\begin{equation}
	E_{\text{post}} = \langle \psi_{\text{post}}|\hat{H}|\psi_{\text{post}} \rangle \approx -h(\langle\hat{\sigma}^z_1\rangle + \langle\hat{\sigma}^z_N\rangle) \approx -2h.
\end{equation}
Consequently, the average energy injected per trial scales linearly with distance:
\begin{equation}
	\Delta E_{\text{inj}} = E_{\text{post}} - E_{\text{pre}} \approx -2h - (-Nh) = (N - 2)h.
		\label{eq:energy_injection_derived}
\end{equation}
Combining the exponential number of runs with the linear energy injection, the total average energy cost becomes:
\begin{equation}
	\langle E_{\text{total}} \rangle = \langle N_{\text{runs}} \rangle \times \Delta E_{\text{inj}} \approx (2^{N-2}) \times ((N-2)h).
	\label{eq:total_cost_derived}
\end{equation}
This result confirms the fatal flaw of the monolithic protocol. The total cost scales as a polynomial multiplied by an exponential ($\sim N \cdot 2^N$). This scaling is visualized in Fig.~\ref{fig:monolithic_cost_analysis}, which breaks down the contributing factors and shows the resulting insurmountable cost wall.

In conclusion, while measurement can induce non-local correlations, the price for doing so via this direct, monolithic approach is an exponentially divergent resource cost. This fundamental scaling barrier motivates the development of a hierarchical, repeater-based architecture, which will be the subject of the next section.
	
	
\section{Protocol II: The Scalable Quantum Repeater Architecture}
\label{sec:repeater_protocol}

To overcome the exponential cost barrier inherent in the monolithic protocol, we now introduce a hierarchical architecture inspired by quantum repeater networks. We demonstrate that by segmenting the problem into manageable steps, it is possible to transform the exponential resource scaling into a polynomial one, thus rendering long-range QET physically feasible. This section breaks down the architecture layer by layer, starting with the basic structure and its immediate advantages.

\subsection{Basic Architecture: Segmentation and Entanglement Swapping}
\label{subsec:segmentation}

The foundational concept of the repeater architecture is to abandon the single, long-distance channel in favor of a chain of shorter, concatenated elementary links.

\begin{figure}[t]
	\centering
	\includegraphics[width=0.8\columnwidth]{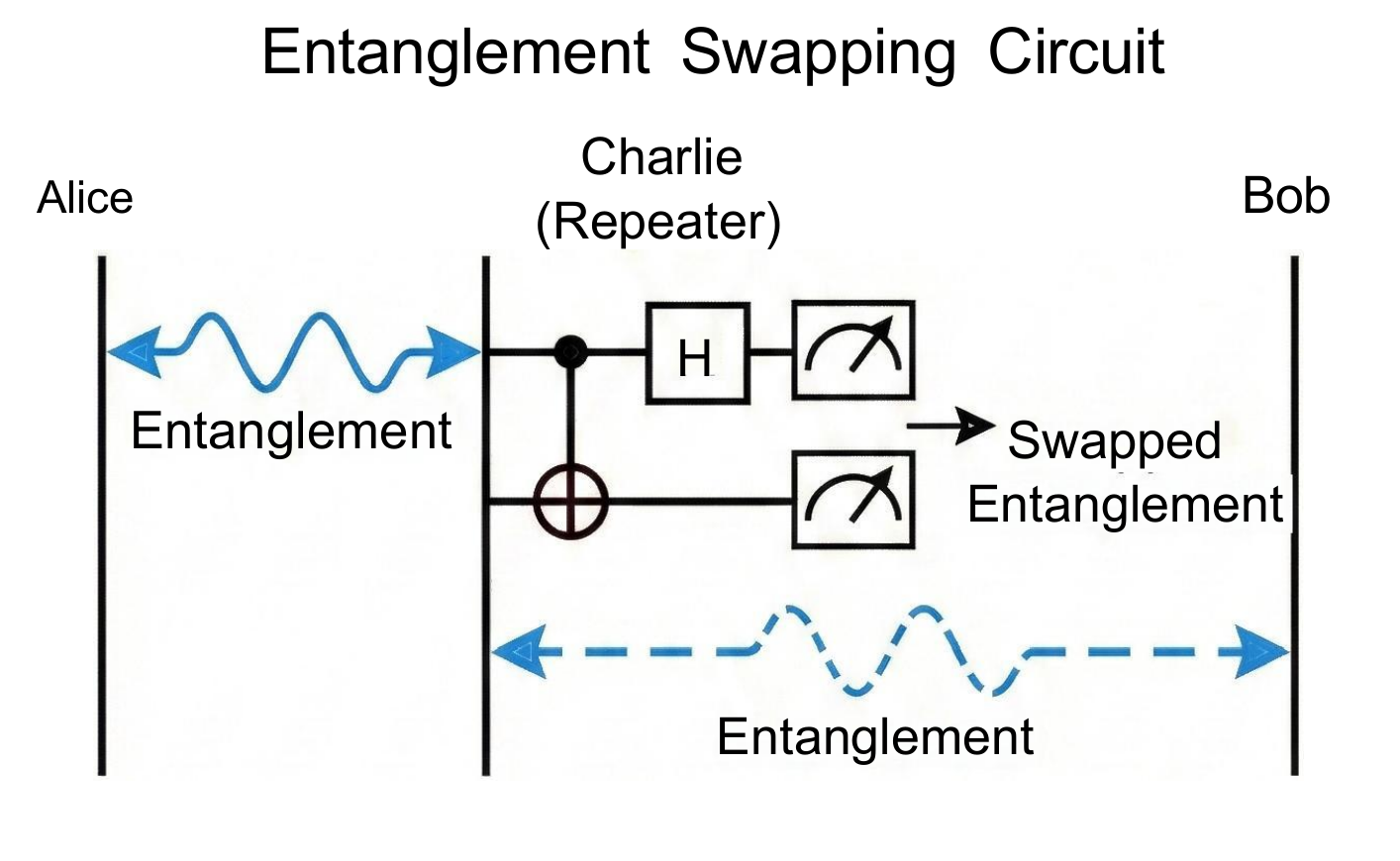}
	\caption{The Entanglement swapping operation. To connect two adjacent entangled links (e.g., Alice-Charlie and Charlie-Bob), a Bell State Measurement (BSM) is performed on the two qubits held by the intermediate repeater station (Charlie). This projects the system into a new state where Alice and Bob are directly entangled, effectively extending the range of the entanglement.}
	\label{fig:entanglement_swapping}
\end{figure}

\begin{enumerate}
	\item  The total distance $N$ is partitioned into $M$ adjacent, non-overlapping segments of a short, constant length $L$, such that $N = ML$. This segmented architecture fundamentally relies on quantum memories at each repeater station, capable of storing the entangled qubits with high fidelity while other elementary links are being established.
	\item Layer 1 - Heralded link generation: entanglement is generated in parallel across all $M$ elementary segments. This is a probabilistic process. For each segment, attempts are made until a successful entangled pair is created and heralded (i.e., its creation is announced).
	\item Layer 2 - Entanglement swapping: Once all $M$ short-range entangled links have been established and stored in quantum memories with coherence times significantly exceeding the total generation time (e.g., between sites $(1,L), (L,2L), \dots, ((M-1)L, ML)$), a Bell State Measurement (BSM) is performed at each intermediate node (sites $L, 2L, \dots, (M-1)L$). As illustrated in Fig.~\ref{fig:entanglement_swapping}, a successful BSM on the qubits at an intermediate node effectively connects the two adjacent links, creating a longer-distance entangled pair between the outer nodes. This process is repeated $M-1$ times along the chain to establish a single, end-to-end entangled channel between Alice (site 1) and Bob (site $N$).
\end{enumerate}

This segmented approach fundamentally alters the temporal scaling of the protocol. In the monolithic case, the waiting time for success scaled exponentially with the total distance $N$. Here, the primary time cost is determined by the slowest of the $M$ parallel link generation processes.

Let $P_L$ be the constant success probability of generating an entangled link over a short segment of length $L$. From our analysis in Sec.~\ref{sec:monolithic_protocol}, $P_L \approx 2^{-(L-2)}$. The problem of waiting until all $M$ segments have succeeded is a classic scenario in probability theory known as the coupon collector's problem. The expected number of parallel rounds, $\langle R \rangle$, required to complete all $M$ links is given by (see Appendix~\ref{app:coupon_collector} for a derivation):
\begin{equation}
	\langle R(M) \rangle = \frac{1}{P_L} \sum_{k=1}^{M} \frac{1}{k} = \frac{H_M}{P_L},
	\label{eq:coupon_collector}
\end{equation}
where $H_M$ is the $M$-th harmonic number. For large $M$, the harmonic number is well-approximated by the natural logarithm, $H_M \approx \ln(M) + \gamma_E$, where $\gamma_E$ is the Euler-Mascheroni constant. Since $M=N/L$, the time cost scales polylogarithmically with the total distance $N$:
\begin{equation}
	\langle R(N) \rangle \approx \frac{\ln(N/L)}{P_L} \in \mathcal{O}(\log N).
	\label{eq:log_scaling_time}
\end{equation}
This result represents a monumental improvement, converting an exponential time barrier into a manageable polylogarithmic one. However, as we will demonstrate in the next subsection, this simple iterative architecture introduces new challenges related to fidelity and the probabilistic nature of entanglement swapping.

\subsection{New Barriers: Fidelity Decay and Probabilistic Swapping}
\label{subsec:new_barriers}
While the segmented architecture fundamentally solves the temporal scaling, this idealized picture conceals critical barriers that arise from operational imperfections. We now analyze these barriers by introducing phenomenological models for noise and probabilistic success, which are standard in the analysis of quantum repeater protocols \cite{RepeaterReview_Sangouard}.

\begin{figure}[t]
	\centering
\includegraphics[width=\columnwidth,height=6cm]{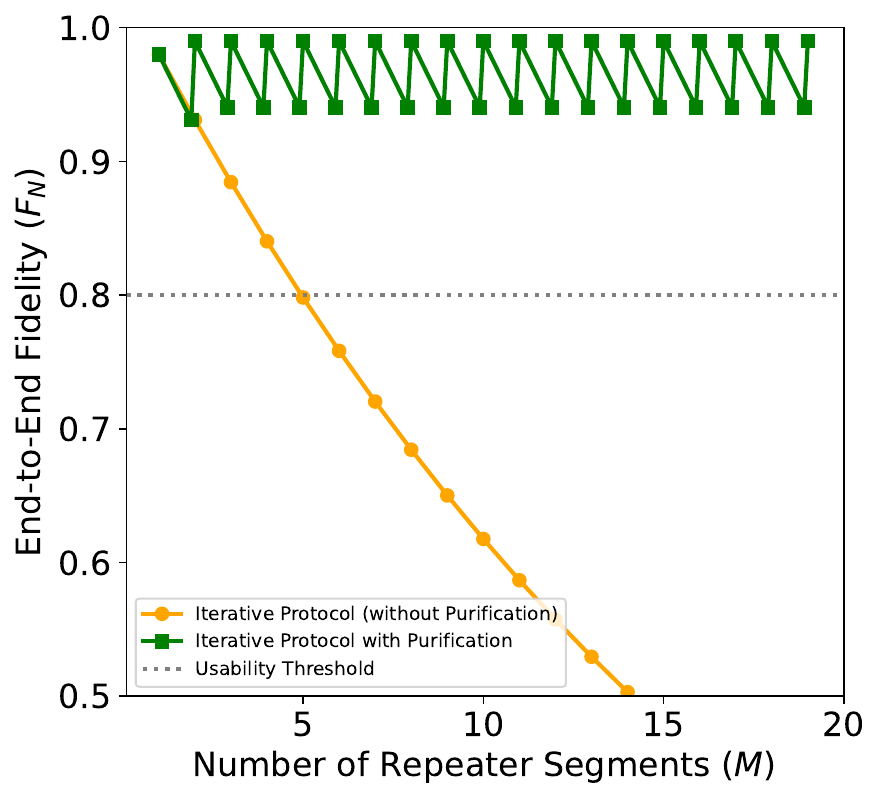}
\caption{The simple iterative protocol (orange circles), governed by the model in Eq.~\eqref{eq:fidelity_decay}, exhibits exponential fidelity decay. The full repeater protocol with purification (green squares, discussed in Sec.~\ref{subsec:purification}) is required to counteract this decay.}
	\label{fig:fidelity_decay_comparison}
\end{figure}

We first analyze the impact of noise on the quality of the entangled state, quantified by the fidelity $F$. We model the noise using two experimentally relevant parameters:
\begin{itemize}
	\item $F_L$: The average initial fidelity of a heralded elementary link over a short segment of length $L$.
	\item $F_{\text{swap}}$: The operational fidelity of the entanglement swapping gate, representing the noise added at each step.
\end{itemize}
To understand the scaling barrier, consider first a linear chain of $M-1$ swapping operations. In this baseline scenario, errors accumulate sequentially, and the final fidelity decays exponentially \cite{NielsenChuang}:
\begin{equation}
	F_N \approx F_L \cdot (F_{\text{swap}})^{M-1} = F_L \cdot (F_{\text{swap}})^{(N/L)-1}.
	\label{eq:fidelity_decay}
\end{equation}
This equation, although a useful model, reveals a critical exponential barrier. For any realistic imperfection ($F_{\text{swap}} < 1$), the fidelity decays exponentially with the number of segments $M$. This behavior is visualized in Fig.~\ref{fig:fidelity_decay_comparison}, where the fidelity rapidly drops below the entanglement threshold, rendering the channel useless.

A second barrier arises from the probabilistic nature of the BSMs. Let $p_{\text{BSM}} < 1$ be the success probability of a single BSM. To establish the end-to-end link, all $M-1$ swapping operations must succeed. These are independent Bernoulli trials, so the total success probability of the swapping chain is the product of the individual probabilities:
\begin{equation}
	P_{\text{swap-chain}} = (p_{\text{BSM}})^{M-1} = (p_{\text{BSM}})^{(N/L)-1}.
	\label{eq:swap_chain_prob}
\end{equation}
For any non-deterministic BSM ($p_{\text{BSM}} < 1$), this probability also decays exponentially with $N$. This reintroduces an exponential cost barrier typical of linear chains. As we show next, overcoming this requires combining entanglement purification with a nested architecture. The total average energy cost is the cost of successfully generating all elementary links, $\langle E_{\text{links}} \rangle$, divided by this success probability. 

The total energy cost to generate the $M$ elementary links, unlike the logarithmic time cost derived in Appendix~\ref{app:coupon_collector}, scales linearly with the number of links. Since each of the $M=N/L$ links has a constant average generation cost (derived from Eq.~\eqref{eq:total_cost_derived} for a fixed small $L$), the total energy cost for this layer scales as $\langle E_{\text{links}} \rangle \in \mathcal{O}(M) = \mathcal{O}(N)$.
The total cost for the simple iterative protocol is therefore:
\begin{equation}
	\langle E_{\text{total}} \rangle_{\text{simple-iter}} = \frac{\langle E_{\text{links}} \rangle}{P_{\text{swap-chain}}} \approx \frac{c \cdot N}{(p_{\text{BSM}})^{(N/L)-1}},
	\label{eq:simple_iter_total_cost}
\end{equation}
where $c \approx (L-2)h$ represents the average energy cost to generate a single elementary link of length $L$. As $(p_{\text{BSM}})^{-N/L} = e^{N/L \cdot \ln(1/p_{\text{BSM}})}$, this cost scales as $\mathcal{O}(N \cdot e^{\alpha N})$. This proves that the simple iterative protocol is not scalable, as it reintroduces an exponential cost factor.

\subsection{The Complete Solution: Entanglement Purification}
\label{subsec:purification}
To overcome the exponential barriers of fidelity decay inherent in linear models, we adopt a hierarchical architecture. By combining nested entanglement swapping (which reduces the effective gate depth from linear to logarithmic) with the critical layer of entanglement purification \cite{Purification_DEJMPS}. Purification is a protocol that allows two parties to distill a single high-fidelity entangled pair from multiple lower-fidelity pairs.

\begin{figure}[h!]
	\centering
	\includegraphics[width=0.7\columnwidth]{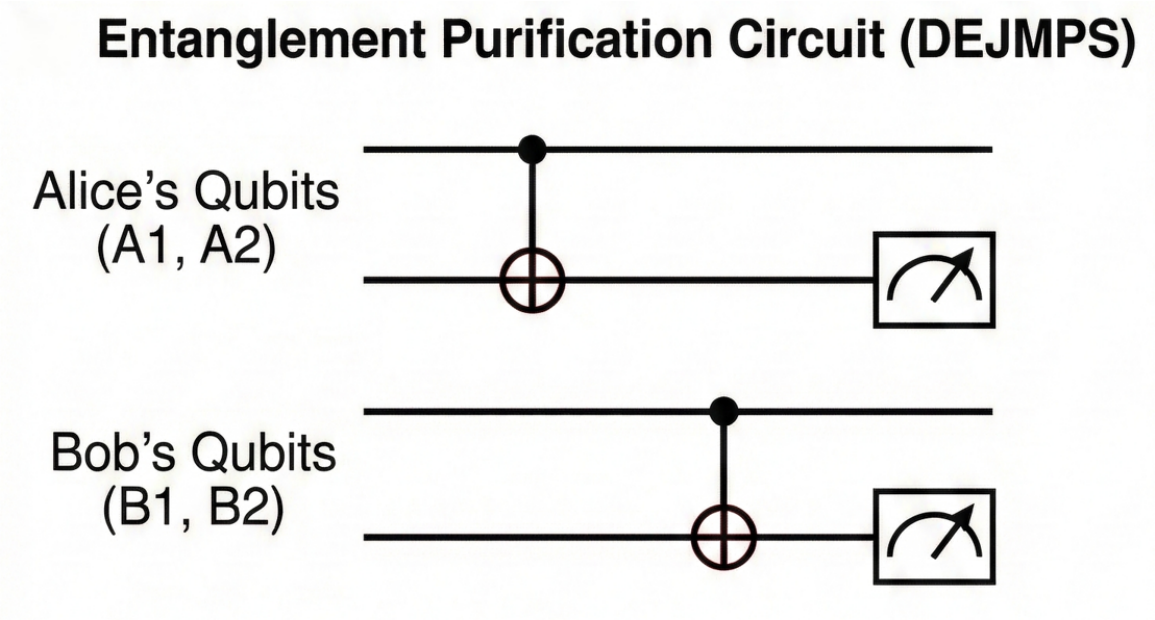}
	\caption{The DEJMPS entanglement purification circuit. Two noisy pairs are consumed. Local CNOTs and measurements are performed. Conditioned on matching outcomes, the remaining pair (Pair 1) is kept with improved fidelity.}
	\label{fig:purification_circuit}
\end{figure}

We illustrate this principle using a protocol introduced by Deutsch--Ekert--Josza--Macchiavello--Popescu--Sanpera (DEJMPS), whose quantum circuit is shown in Fig.~\ref{fig:purification_circuit}. The protocol proceeds via the following steps:

\begin{enumerate}
	\item Alice and Bob begin with two independent noisy entangled pairs, labeled Pair 1 (qubits $A_1, B_1$) and Pair 2 (qubits $A_2, B_2$). Each pair has the same initial fidelity $F$.
	
	\item  Alice performs a CNOT gate on her two qubits, with $A_1$ as control and $A_2$ as target ($\text{CNOT}_{A_1 \to A_2}$). Simultaneously, Bob performs a similar CNOT gate on his qubits ($\text{CNOT}_{B_1 \to B_2}$).
	
	\item Alice measures her second qubit ($A_2$) in the computational basis $\hat{\sigma}^z$, obtaining a classical outcome $m_A \in \{0, 1\}$. Bob does the same for his qubit $B_2$, obtaining outcome $m_B$.
	
	\item  Alice and Bob communicate their classical outcomes. If the outcomes match ($m_A = m_B$), the protocol is successful. They keep Pair 1, which now possesses an enhanced fidelity $F'$, and discard Pair 2. If the outcomes do not match, the protocol fails, and both pairs are discarded.
\end{enumerate}

To analyze this protocol's performance, we model the noisy pairs as Bell-diagonal states, a general model for errors in quantum channels. As rigorously derived in Appendix~\ref{app:purification_derivation}, a single round of purification asymmetrically suppresses different types of errors. To achieve optimal performance, the full repeater architecture employs the iterative DEJMPS method, which includes local basis permutations (e.g., local rotations) between subsequent rounds. This strategy ensures that all error types are suppressed effectively \cite{Purification_DEJMPS}. The crucial property of this iterative process is its behavior in the high-fidelity limit. By defining the error as $\epsilon = 1 - F$, the analysis in Appendix~\ref{app:purification_derivation} shows that a full cycle of the purification protocol (e.g., two rounds with an intermediate basis rotation) transforms the error according to the map:
\begin{equation}
	\epsilon' \approx \mathcal{O}\epsilon^2.
	\label{eq:quadratic_convergence}
\end{equation}
This quadratic convergence is the central mathematical result. It allows the definition of a target fidelity, $F_{\text{target}}$. At each repeater node, purification rounds are performed until the link fidelity exceeds this threshold. This procedure replaces the decaying fidelity from Eq.~\eqref{eq:fidelity_decay} with a periodic reset, as shown in Fig.~\ref{fig:fidelity_decay_comparison}, ensuring a scalable protocol.

\subsection{
Comparative Verdict}
\label{subsec:final_analysis}
\begin{figure}[h!]
	\centering
\includegraphics[width=\columnwidth,height=5cm]{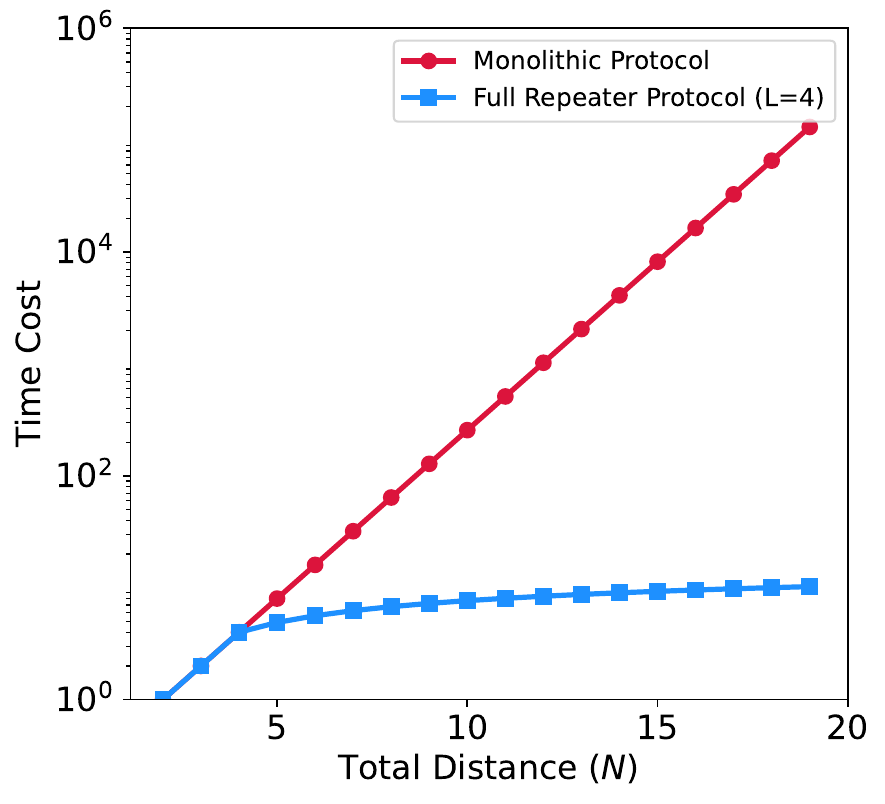}\put(-195,130){$ (a) $}\\
\includegraphics[width=\columnwidth,height=5cm]{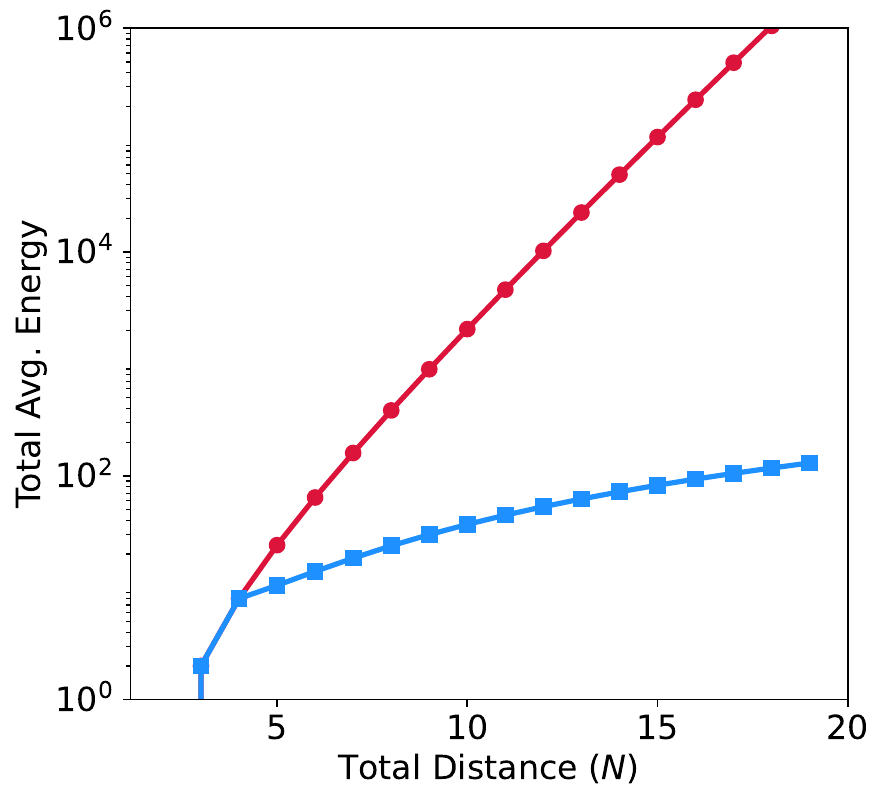}\put(-195,130){$ (b) $}
\caption{ The plots show the (a) time cost and (b) total average energy cost for the monolithic (red) and full repeater (blue) protocols. The repeater architecture converts the exponential scaling of the monolithic approach into manageable polylogarithmic (for time) and polynomial (for energy) scaling.}
	\label{fig:cost_comparison}
\end{figure}

\begin{figure}[h!]
	\centering
\includegraphics[width=\columnwidth,height=5cm]{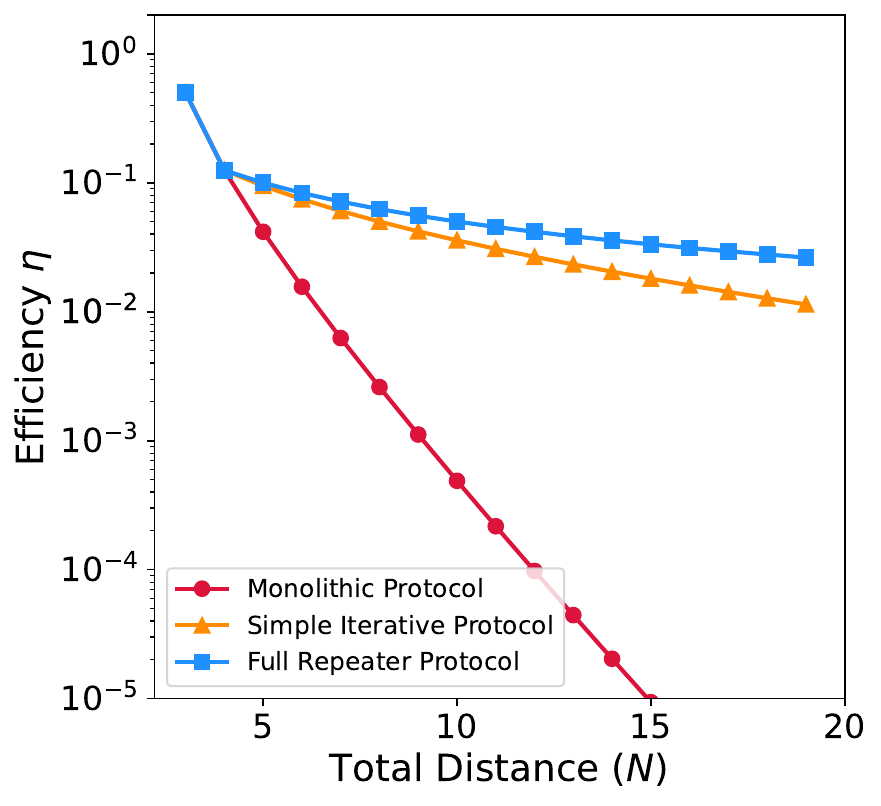}\put(-190,130){$ (a) $}\\
\includegraphics[width=\columnwidth,height=5cm]{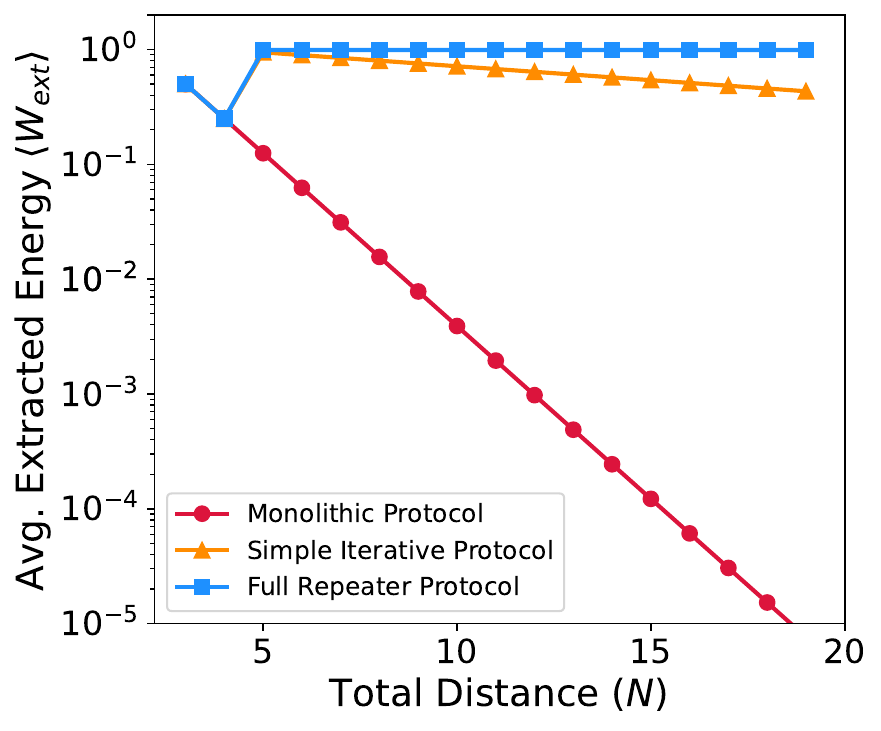}\put(-190,130){$ (b) $} 
\caption{Final verdict on efficiency and payoff.
	(a) Thermodynamic Efficiency ($\eta$): The efficiencies for both the monolithic (red circles) and the simple iterative (orange triangles) protocols decay exponentially to zero, rendering them completely inefficient for long distances. In contrast, the full repeater protocol (blue squares) exhibits a much slower, polynomially decaying efficiency ($\sim 1/\text{poly}(N)$), maintaining a manageable efficiency profile.
(b) Average Extractable Energy ($\langle W_{\text{ext}} \rangle$): The final payoff collapses to zero exponentially for both the monolithic (red) and simple iterative (orange) protocols, due to their vanishing total success probability (either from low probability or low fidelity). The full repeater protocol (blue squares) is the only architecture that overcomes these barriers, maintaining a constant, non-zero average energy yield over large distances and proving its unique scalability.}
	\label{fig:payoff_comparison}
\end{figure}

Having established the complete, multi-layered architecture of the scalable repeater protocol, we now present a final, quantitative comparison between the three strategies discussed: (i) the standard protocol (relying on natural correlations), (ii) the monolithic protocol, and (iii) the full repeater protocol. We analyze the scaling of all critical resources to provide a definitive verdict on scalability. The results are summarized in Fig.~\ref{fig:cost_comparison} and Fig.~\ref{fig:payoff_comparison}.

Figure~\ref{fig:cost_comparison} provides a direct comparison of the primary resource costs. Figure~\ref{fig:cost_comparison}(a) shows the time cost (average rounds). The monolithic protocol's time cost (red circles) scales exponentially ($\sim 2^N$) due to its reliance on a single probabilistic event over the entire distance. In stark contrast, the full repeater protocol (blue squares), which leverages parallelization in Layer 1, exhibits a polylogarithmic scaling ($\sim \log N$), representing an exponential speedup. Figure~\ref{fig:cost_comparison}(b) illustrates the total average energy cost. Here, both the monolithic (red) and the simple iterative (orange, not shown for clarity but scales similarly to red due to probabilistic swapping) protocols exhibit an exponential cost wall ($\sim N \cdot 2^N$). The full repeater protocol (blue), however, which replaces exponential probabilities with polynomial overheads from purification, scales polynomially ($\sim \text{poly}(N)$). This demonstrates its physical feasibility for large distances.

Although operational feasibility is crucial, the ultimate measure of a QET protocol is its ability to deliver a non-vanishing result. Figure~\ref{fig:payoff_comparison} analyzes the final output.
Figure~\ref{fig:payoff_comparison}(a) shows the thermodynamic efficiency ($\eta$). While the efficiency of all protocols decays with distance, the decay for the monolithic and simple iterative protocols is exponential, rendering them useless. The full repeater's efficiency decays only polynomially ($\sim 1/\text{poly}(N)$), maintaining a manageable, non-zero efficiency profile.

Figure~\ref{fig:payoff_comparison} (b) presents the most important metric: the average extractable energy, $\langle W_{\text{ext}} \rangle = h \cdot P_{\text{total-succ}}$. Both the monolithic (red) and simple iterative (orange) protocols have an average yield that collapses to zero exponentially, due to vanishing success probability and/or fidelity. The full repeater protocol (blue) is the only architecture that maintains a constant, non-zero average energy payoff, independent of distance $N$ for large $N$. Crucially, a local repeat-until-success strategy bypasses the probabilistic decay of Eq. (\ref{eq:swap_chain_prob}), converting exponential failure rates into manageable polynomial time overheads, thereby ensuring the near-deterministic yield.

In conclusion, the analysis confirms that only the full repeater architecture, which systematically addresses the exponential barriers of time, fidelity, and probability, provides a truly scalable and physically viable blueprint for long-range QET. The scaling laws are summarized in Table~\ref{tab:scaling_summary}.

\begin{table}[h!]
	\centering
	\caption{Summary of asymptotic scaling laws for QET protocols. Here, $N$ denotes the total distance. The exponents $\alpha, \beta,$ and $\kappa$ represent positive constants determined by physical parameters, $\beta \sim \xi^{-1}$ is governed by the correlation length, while $\alpha$ and $\kappa$ depend on the repeater link length $L$ and the swapping success probability $p_{\text{BSM}}$.}
	\label{tab:scaling_summary}
    \resizebox{\columnwidth}{!}{
	\begin{tabular}{lccc}
		\hline\hline
		\textbf{Metric} & \textbf{Monolithic} & \textbf{Simple Iterative} & \textbf{Full Repeater} \\
		\hline
		Time Cost $\langle R \rangle$ & $\mathcal{O}(2^N)$ & $\mathcal{O}(\log N)$ & $\mathcal{O}(\text{poly}(\log N))$ \\
		Energy Cost $\langle E \rangle$ & $\mathcal{O}(N \cdot 2^N)$ & $\mathcal{O}(e^{\alpha N})$ & $\mathcal{O}(\text{poly}(N))$ \\
		Efficiency $\eta$ & $\mathcal{O}(e^{-\beta N})$ & $\mathcal{O}(e^{-\kappa N})$ & $\mathcal{O}(1/\text{poly}(N))$ \\
		Final Yield $\langle W \rangle$ & $\to 0$ (exp.) & $\to 0$ (exp.) & Constant \\
		\hline\hline
	\end{tabular}}
\end{table}

\section{The Complete Scalable QET Protocol and Final Implementation}
\label{sec:final_protocol}

\begin{figure}[b]
	\centering
	\includegraphics[width=0.8\columnwidth]{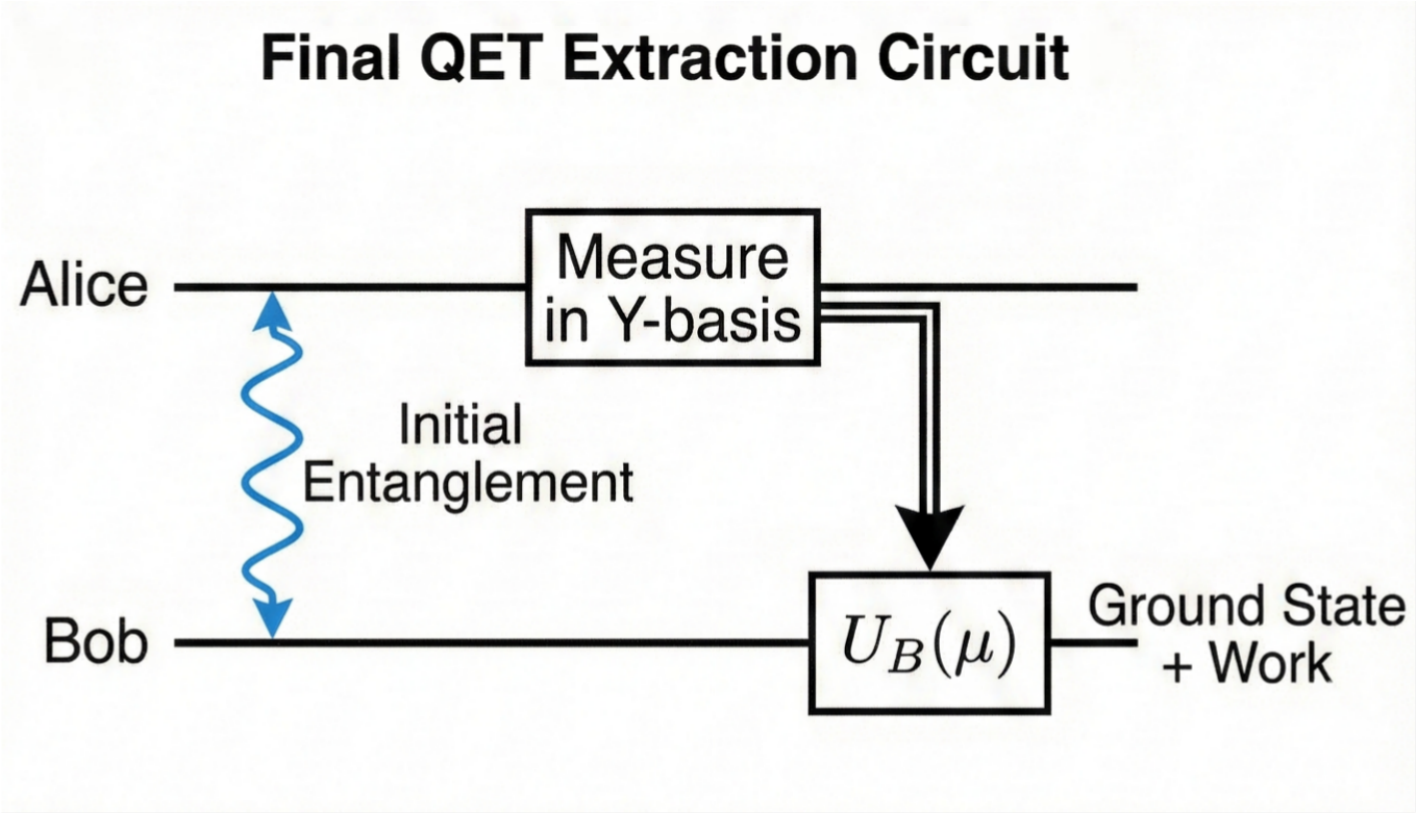}
	\caption{The final QET extraction circuit. Alice performs a local measurement and communicates the result $\mu$ to Bob, who applies a conditional unitary $\hat{U}_B(\mu)$ to harvest energy.}
	\label{fig:final_qet_circuit}
\end{figure}

Having established the scalable repeater architecture in Sec.~\ref{sec:repeater_protocol}, we now detail the final steps for energy extraction. The purification process ensures that Alice ($A$) and Bob ($B$) share a high-fidelity entangled link. To precisely define the energy extraction, we first specify the local Hamiltonian for Bob's qubit as $H_B = -h\sigma_z^B$. In this convention, the spin-up state $|0\rangle_B$ is the ground state with energy $E_{\text{ground}} = -h$. The extraction protocol, illustrated in Fig.~\ref{fig:final_qet_circuit}, proceeds as follows:

\begin{enumerate}
	\item  Alice performs a local projective measurement on her qubit in the $\hat{\sigma}_y$ basis. Although correlations are $xx$-dominant, a transverse measurement is required to induce non-commuting fluctuations, steering Bob’s state into an active superposition capable of yielding energy.
	\item Alice communicates her measurement outcome $\mu \in \{+1, -1\}$ to Bob via a classical channel.
	\item  Conditioned on the value of $\mu$, Bob applies a specific local unitary operation $\hat{U}_B(\mu)$ to his qubit to extract work.
\end{enumerate}

The entanglement purification layer (Sec.~\ref{subsec:purification}) drives the fidelity of the shared link towards unity. We are therefore justified in treating the shared channel as an ideal maximally entangled Bell state:
\begin{equation}
	|\Phi^+\rangle_{AB} = \frac{1}{\sqrt{2}}(|0\rangle_A|0\rangle_B + |1\rangle_A|1\rangle_B).
\end{equation}
Before Alice's measurement, Bob's local state is obtained by tracing out Alice's system:
\begin{equation}
	\hat{\rho}_B = \text{Tr}_A(|\Phi^+\rangle\langle\Phi^+|) = \frac{1}{2}\mathbf{I}.
\end{equation}
This maximally mixed state is passive with respect to Bob's local Hamiltonian $\hat{H}_B = -h\hat{\sigma}_z^B$; no energy can be extracted from it by unitary operations.

Alice measures the observable $\hat{\sigma}_y^A$. The measurement projectors correspond to outcomes $\mu = \pm 1$:
\begin{equation}
	 \hat{P}_\mu^A = \frac{1}{2}(\mathbf{I} + \mu\hat{\sigma}_y^A).
\end{equation}
Upon obtaining outcome $\mu$, the global state collapses, and Bob's state is updated to the conditional pure state $|\psi_B(\mu)\rangle$:
\begin{align}
	|\psi_B(\mu)\rangle &= \frac{\left(\langle \mu_y |_A \otimes I_B\right).|\Phi^+\rangle_{AB}}{ \sqrt{p(\mu)}}  \nonumber \\
	&= \frac{1}{\sqrt{2}} (|0\rangle_B - i\mu|1\rangle_B),
	\label{eq:bob_conditional_state}
\end{align}
where $ p(\mu) = \text{Tr}\left( \hat{P}^A_{\mu} \hat{\rho}_A \right)=1/2 $, and  $| \mu_y \rangle$  represents the eigenstate of the Pauli-Y operator corresponding to the measurement outcome $\mu$. This state is a superposition of energy eigenstates and is therefore active.

We calculate the ergotropy $\mathcal{E}$, defined as the difference between the state's mean energy and the ground state energy. In the paramagnetic phase ($h > 1$), the local energy is dominated by the interaction with the external field. Therefore, defining the effective local Hamiltonian by the dominant transverse field term $\hat{H}_B = -h\hat{\sigma}_z^B$, the initial energy of Bob's conditional state is:
\begin{align}
	E_{\text{initial}} &= \langle \psi_B(\mu) | \hat{H}_B | \psi_B(\mu) \rangle \nonumber \\
	&= \frac{1}{2} ({}_B\langle 0| + i\mu \ {}_B\langle 1|) (-h\hat{\sigma}_z^B) (|0\rangle_B - i\mu|1\rangle_B) \nonumber \\
	&= 0.
\end{align}
The ground state of $\hat{H}_B = -h\hat{\sigma}_z^B$ is $|0\rangle$, with energy $E_{\text{ground}} = -h$. The extractable work is thus:
\begin{equation}
	\mathcal{E}(\psi_B) = E_{\text{initial}} - E_{\text{ground}} = 0 - (-h) = h.
	\label{eq:ergotropy_result}
\end{equation}
This result rigorously proves that a finite quantum of energy $h$ becomes available for extraction.

Bob extracts this energy by applying a unitary $\hat{U}_B(\mu)$ that rotates $|\psi_B(\mu)\rangle$ to the ground state $|0\rangle_B$. This rotation is described by:
\begin{equation}
	\hat{U}_B(\mu) = \exp\left(-i \mu \frac{\pi}{4} \sigma_x^B\right) = \frac{1}{\sqrt{2}} \left( I - i\mu\sigma_x^B \right).
\end{equation}
Applying this unitary to Bob's conditional state confirms the successful extraction:
\begin{equation}
	\hat{U}_B(\mu)|\psi_B(\mu)\rangle = \frac{1}{2}\left( I - i\mu\sigma_x^B \right) \left( |0\rangle_B - i\mu|1\rangle_B \right) = |0\rangle_B.
\end{equation}
By reaching the ground state, Bob harvests the energy difference $W = h$. This concludes the protocol, demonstrating that energy teleportation is physically realizable over arbitrary distances using our scalable architecture.

\section{Conclusion and Outlook}
\label{sec:conclusion}

In this work, we have addressed the fundamental challenge of exponential suppression of entanglement and extractable energy over macroscopic distances in gapped many-body systems. By systematically analyzing the thermodynamic limitations of measurement-induced protocols, we have demonstrated that overcoming the locality of the vacuum requires a scalable architecture capable of fault-tolerant entanglement distribution.

Our analysis rigorously established that while the correlation length $\xi$ in the XY model dictates a severe exponential decay $e^{-N/\xi}$, this limit can be bypassed via non-local measurements. However, we proved that a direct monolithic strategy is physically untenable; despite establishing a non-local link, the resource cost required to achieve a successful outcome scales exponentially as $\mathcal{O}(2^N)$, creating an insurmountable barrier for large systems. The primary contribution of this work is the resolution of this scaling crisis through a hierarchical quantum repeater architecture. Our protocol guarantees a constant, non-vanishing yield of activated energy $\langle W \rangle = h$ at any distance, albeit at a net thermodynamic cost. This result reframes long-range QET from a theoretical curiosity into a tangible protocol for remote quantum resource activation.

These results shift the perspective on long-range QET from a theoretical curiosity constrained by vanishing probabilities to a tangible resource optimization problem. The teleported energy is demystified as the consumption of a pre-established, high-fidelity entanglement resource whose cost is manageable. Future work may focus on optimizing purification protocols to further reduce the polynomial overhead, or extending this formalism to higher-dimensional systems. Ultimately, we have provided a viable blueprint for scalable QET. This demonstrates that the vacuum can serve as a medium for remotely triggering local quantum processes, such as activating topological qubits or initiating quantum computations, provided the necessary quantum infrastructure is established.

\appendix
		
	\section{Derivation of the Polylogarithmic Time Cost}
	\label{app:coupon_collector}
	
	This appendix provides the mathematical derivation for the polylogarithmic scaling of the heralded link generation time, as cited in Eq.~\eqref{eq:log_scaling_time}. The problem is a direct application of the classic coupon collector's problem.
	
	Let $R$ be the total number of parallel rounds required to establish all $M$ elementary links. We can express $R$ as a sum of random variables:
	\begin{equation}
		R = T_1 + T_2 + \dots + T_M,
	\end{equation}
	where $T_k$ is the number of additional rounds required to secure the $k$-th new successful link, given that $k-1$ links have already been established.
	
	At the stage where $k-1$ links are complete, there are $M-(k-1)$ segments still attempting to generate entanglement. The probability that any single one of these remaining segments succeeds in a given round is $P_L$. The probability that it fails is $(1-P_L)$. The probability that all of the remaining $M-(k-1)$ segments fail in one round is:
	\begin{equation}
		P(\text{all fail}) = (1 - P_L)^{M-k+1}.
	\end{equation}
	Therefore, the probability of obtaining new successful link in this round, which we denote $p_k$, is:
	\begin{equation}
		p_k = 1 - (1 - P_L)^{M-k+1}.
	\end{equation}
	The random variable $T_k$ follows a geometric distribution with success probability $p_k$. Its expected value is the inverse of this probability:
	\begin{equation}
		\mathbb{E}[T_k] = \frac{1}{p_k} = \frac{1}{1 - (1 - P_L)^{M-k+1}}.
	\end{equation}
	By the linearity of expectation, the total average number of rounds is the sum of these expected values:
	\begin{equation}
		\langle R \rangle = \mathbb{E}[R] = \sum_{k=1}^{M} \mathbb{E}[T_k] = \sum_{k=1}^{M} \frac{1}{1 - (1 - P_L)^{M-k+1}}.
	\end{equation}
	To find the asymptotic scaling for large $M$, we analyze the expression for $p_k$. Since $L$ is a small constant, $P_L = 2^{-(L-2)}$ is also a small constant. For small $x$, we can use the binomial approximation $(1-x)^n \approx 1-nx$. Let $j = M-k+1$ be the number of remaining links to be collected. Then:
	\begin{equation}
		p_k \approx 1 - (1 - j \cdot P_L) = j \cdot P_L.
	\end{equation}
	Substituting this approximation back into the sum and changing the summation index from $k$ to $j$ (where $j$ runs from $M$ down to $1$):
	\begin{equation}
		\langle R \rangle \approx \sum_{j=1}^{M} \frac{1}{j \cdot P_L} = \frac{1}{P_L} \sum_{j=1}^{M} \frac{1}{j}.
	\end{equation}
	The sum $\sum_{j=1}^{M} \frac{1}{j}$ is the $M$-th harmonic number, $H_M$.
	\begin{equation}
		\langle R \rangle \approx \frac{H_M}{P_L}.
	\end{equation}
	For large $M$, the harmonic number is excellently approximated by the natural logarithm:
	\begin{equation}
		H_M \approx \ln(M) + \gamma_E + \mathcal{O}(1/M),
	\end{equation}
	where $\gamma_E \approx 0.577$ is the Euler-Mascheroni constant. Substituting $M=N/L$ and the constant value of $P_L=2^{-(L-2)}$, we obtain the final asymptotic scaling law for the time cost:
	\begin{equation}
		\langle R(N) \rangle \approx \frac{\ln(N/L) + \gamma_E}{2^{-(L-2)}} \in \mathcal{O}(\log N).
	\end{equation}
	This rigorously demonstrates the polylogarithmic scaling of the time cost.

\section{Derivation of the Fidelity Transformation Map}
\label{app:purification_derivation}

We begin with the Werner state $\hat{\rho}(F)$, which is a mixture of the pure Bell state $|\Phi^+\rangle$ and white noise (an equal mixture of the four Bell states). It is defined by its fidelity $F$ with respect to $|\Phi^+\rangle$:
\begin{equation}
	\hat{\rho}(F) = F |\Phi^+\rangle\langle\Phi^+| + \frac{1-F}{3} \sum_{i \in \{\Psi^+, \Phi^-, \Psi^-\}} |\Psi_i\rangle\langle\Psi_i|.
	\label{eq:werner_state}
\end{equation}
This state is a specific type of a more general class called Bell-diagonal states. A Bell-diagonal state is fully described by its four diagonal elements in the Bell basis $\{|\Phi^+\rangle, |\Phi^-\rangle, |\Psi^+\rangle, |\Psi^-\rangle\}$, which we denote as $(A, B, C, D)$ respectively.

By comparing Eq.~\eqref{eq:werner_state} with the general form, one can directly map the fidelity $F$ to these components:
\begin{align}
	A &= \langle\Phi^+|\hat{\rho}(F)|\Phi^+\rangle = F, \\
	B &= \langle\Phi^-|\hat{\rho}(F)|\Phi^-\rangle = \frac{1-F}{3}, \\
	C &= \langle\Psi^+|\hat{\rho}(F)|\Psi^+\rangle = \frac{1-F}{3}, \\
	D &= \langle\Psi^-|\hat{\rho}(F)|\Psi^-\rangle = \frac{1-F}{3}.
\end{align}
This explicitly shows that for a Werner state, the error components are symmetric: $B=C=D$.
We define the total error as $\epsilon = 1-F$. Using this, we can express the initial components in terms of $\epsilon$:
\begin{equation}
	A = 1 - \epsilon, \quad B = C = D = \frac{\epsilon}{3}.
	\label{eq:werner_errors}
\end{equation}
This is the starting point for our analysis.

The CNOT-based purification protocol (DEJMPS) operates on two identical copies of a Bell-diagonal state. If the protocol succeeds, the output state $\rho'$ has new diagonal elements $(A', B', C', D')$ given by the map:
\begin{align}
	&A' = \frac{A^2 + B^2}{N}, \quad
	B' = \frac{2CD}{N}, \nonumber \\
	&
	C' = \frac{C^2 + D^2}{N}, \quad
	D' = \frac{2AB}{N},
	\label{eq:map}
\end{align}
where $N = (A+B)^2 + (C+D)^2$ is the success probability.

We now apply the map to our initial Werner state components from Eq. \eqref{eq:werner_errors} and perform a Taylor expansion for small $\epsilon \ll 1$, namely
\begin{align}
	N_1 &= (A+B)^2 + (C+D)^2 \\
	&= \left((1-\epsilon) + \frac{\epsilon}{3}\right)^2 + \left(\frac{\epsilon}{3} + \frac{\epsilon}{3}\right)^2 \\
	&= \left(1 - \frac{2\epsilon}{3}\right)^2 + \left(\frac{2\epsilon}{3}\right)^2 \\
	&= \left(1 - \frac{4\epsilon}{3} + \frac{4\epsilon^2}{9}\right) + \frac{4\epsilon^2}{9} = 1 - \frac{4\epsilon}{3} + \frac{8\epsilon^2}{9}.
\end{align}

Let's denote the components after the first round as $(A_1, B_1, C_1, D_1)$. We calculate them keeping terms up to $\mathcal{O}(\epsilon^2)$, given by
\begin{align}
	B_1 &= \frac{2CD}{N_1} = \frac{2(\epsilon/3)(\epsilon/3)}{1 - \frac{4\epsilon}{3} + \mathcal{O}(\epsilon^2)}\nonumber \\
	& = \frac{2\epsilon^2}{9} \left(1 + \frac{4\epsilon}{3} + \dots\right) = \frac{2}{9}\epsilon^2 + \mathcal{O}(\epsilon^3), \\
	C_1 &= \frac{C^2+D^2}{N_1} = \frac{(\epsilon/3)^2 + (\epsilon/3)^2}{1 - \frac{4\epsilon}{3} + \mathcal{O}(\epsilon^2)} \nonumber \\
	&= \frac{2\epsilon^2}{9} \left(1 + \dots\right) = \frac{2}{9}\epsilon^2 + \mathcal{O}(\epsilon^3), \\
	D_1 &= \frac{2AB}{N_1} = \frac{2(1-\epsilon)(\epsilon/3)}{1 - \frac{4\epsilon}{3} + \mathcal{O}(\epsilon^2)} = \frac{\frac{2\epsilon}{3} - \frac{2\epsilon^2}{3}}{1 - \frac{4\epsilon}{3} + \mathcal{O}(\epsilon^2)}\nonumber \\
	&= \left(\frac{2\epsilon}{3} - \frac{2\epsilon^2}{3}\right) \left(1 + \frac{4\epsilon}{3} + \mathcal{O}(\epsilon^2)\right) \nonumber \\
	&= \frac{2\epsilon}{3} + \frac{8\epsilon^2}{9} - \frac{2\epsilon^2}{3} + \mathcal{O}(\epsilon^3) \nonumber \\
	&= \frac{2}{3}\epsilon + \frac{2}{9}\epsilon^2 + \mathcal{O}(\epsilon^3).
\end{align}
The total error after one round, $\epsilon_1 = 1 - A_1 = B_1 + C_1 + D_1$, is dominated by the linear term in $D_1$:
\begin{equation}
	\begin{split}
		\epsilon_1 &= \left(\frac{2}{9}\epsilon^2\right) + \left(\frac{2}{9}\epsilon^2\right) + \left(\frac{2}{3}\epsilon + \frac{2}{9}\epsilon^2\right) + \mathcal{O}(\epsilon^3) \\&= \frac{2}{3}\epsilon + \frac{2}{3}\epsilon^2 + \mathcal{O}(\epsilon^3).
	\end{split}
\end{equation}
Conclusion of Round 1: The error is reduced, but only linearly. The output state is no longer a Werner state, as the errors are now asymmetric: $D_1 \gg B_1, C_1$.

To achieve quadratic convergence, a second purification round is performed on the output state, but after a local basis rotation. A standard choice is a local unitary on one qubit (e.g., a $\pi/2$ Y-rotation) that swaps the roles of the $C$ and $D$ error components.

Let the state before the rotation have components $(A_1, B_1, C_1, D_1)$. After the rotation, the new components $(A_2, B_2, C_2, D_2)$ are:
\begin{align}
	&A_2 = A_1, \quad B_2 = B_1 \approx \frac{2}{9}\epsilon^2, \nonumber\\ &C_2 = D_1 \approx \frac{2}{3}\epsilon, \quad D_2 = C_1 \approx \frac{2}{9}\epsilon^2.
\end{align}
Now, we apply the purification map (see Eq.~\eqref{eq:map}) again to these components. Let the final error components be $(B_3, C_3, D_3)$. The success probability $N_2$ is approximately $1 - \frac{4}{3}\epsilon + \mathcal{O}(\epsilon^2)$.

The key is that the quadratically suppressing terms in the map now act on the largest error component from the previous step, i.e.
\begin{align}
	B_3 &= \frac{2 C_2 D_2}{N_2} = \frac{2 \left(\frac{2}{3}\epsilon + \mathcal{O}(\epsilon^2)\right) \left(\frac{2}{9}\epsilon^2\right)}{1 + \mathcal{O}(\epsilon)} = \frac{8}{27}\epsilon^3 + \mathcal{O}(\epsilon^4), \\
	C_3 &= \frac{C_2^2 + D_2^2}{N_2} = \frac{\left(\frac{2}{3}\epsilon + \mathcal{O}(\epsilon^2)\right)^2 + \left(\frac{2}{9}\epsilon^2\right)^2}{1 + \mathcal{O}(\epsilon)} = \frac{4}{9}\epsilon^2 + \mathcal{O}(\epsilon^3), \\
	D_3 &= \frac{2 A_2 B_2}{N_2} = \frac{2 (1-\mathcal{O}(\epsilon)) (\frac{2}{9}\epsilon^2)}{1 + \mathcal{O}(\epsilon)} = \frac{4}{9}\epsilon^2 + \mathcal{O}(\epsilon^3).
\end{align}
The total error after the full cycle, $\epsilon_{\text{final}} = B_3 + C_3 + D_3$, is now dominated by the quadratic terms:
\begin{equation}
	\epsilon_{\text{final}} = (0) + \left( \frac{4}{9}\epsilon^2 \right) + \left( \frac{4}{9}\epsilon^2 \right) + \mathcal{O}(\epsilon^3) = \frac{8}{9}\epsilon^2 + \mathcal{O}(\epsilon^3).
\end{equation}

After a full cycle consisting of two purification rounds with an intermediate basis rotation, the final error $\epsilon_{\text{final}}$ scales quadratically with the initial error $\epsilon_{\text{initial}}$.
\begin{equation}
	\epsilon_{\text{final}} \approx \frac{8}{9} \epsilon_{\text{initial}}^2.
\end{equation}
This is the final result cited in Eq.~\eqref{eq:quadratic_convergence} of the main text.

\bibliography{bm}

@article{Purification_DEJMPS,
	title = {Quantum privacy amplification and the security of quantum cryptography over noisy channels},
	author = {Deutsch, David and Ekert, Artur and Jozsa, Richard and Macchiavello, Chiara and Popescu, Sandu and Sanpera, Anna},
	journal = {Phys. Rev. Lett.},
	volume = {77},
	issue = {13},
	pages = {2818--2821},
	numpages = {4},
	year = {1996},
	month = {Sep},
	publisher = {American Physical Society},
	doi = {10.1103/PhysRevLett.77.2818}
}

@article{lieb1961two,
	title={Two soluble models of an antiferromagnetic chain},
	author={Lieb, Elliott and Schultz, Theodore and Mattis, Daniel},
	journal={Ann. Phys.},
	volume={16},
	number={3},
	pages={407--466},
	year={1961},
	publisher={Elsevier}
}

@book{bottcher2006analysis,
	title={Analysis of {T}oeplitz operators},
	author={B{\"o}ttcher, Albrecht and Silbermann, Bernd},
	year={2006},
	publisher={Springer}
}

@article{PhysRevA.4.2331,
	title = {Statistical Mechanics of the $\mathrm{XY}$ Model. IV. Time-Dependent Spin-Correlation Functions},
	author = {McCoy, Barry M. and Barouch, Eytan and Abraham, Douglas B.},
	journal = {Phys. Rev. A},
	volume = {4},
	issue = {6},
	pages = {2331--2341},
	numpages = {0},
	year = {1971},
	month = {Dec},
	publisher = {American Physical Society},
	doi = {10.1103/PhysRevA.4.2331},
	url = {https://link.aps.org/doi/10.1103/PhysRevA.4.2331}
}

@article{RevModPhys.82.277,
	title = {Colloquium: Area laws for the entanglement entropy},
	author = {Eisert, J. and Cramer, M. and Plenio, M. B.},
	journal = {Rev. Mod. Phys.},
	volume = {82},
	issue = {1},
	pages = {277--306},
	numpages = {0},
	year = {2010},
	publisher = {American Physical Society},
	doi = {10.1103/RevModPhys.82.277},
	url = {https://link.aps.org/doi/10.1103/RevModPhys.82.277}
}

@article{PhysRevE.85.061126,
	title = {Quantum refrigerators and the third law of thermodynamics},
	author = {Levy, Amikam and Alicki, Robert and Kosloff, Ronnie},
	journal = {Phys. Rev. E},
	volume = {85},
	issue = {6},
	pages = {061126},
	numpages = {9},
	year = {2012},
	publisher = {American Physical Society},
	doi = {10.1103/PhysRevE.85.061126},
	url = {https://link.aps.org/doi/10.1103/PhysRevE.85.061126}
}

@article{RevModPhys.84.621,
	title = {Gaussian quantum information},
	author = {Weedbrook, Christian and Pirandola, Stefano and Garc\'{\i}a-Patr\'on, Ra\'ul and Cerf, Nicolas J. and Ralph, Timothy C. and Shapiro, Jeffrey H. and Lloyd, Seth},
	journal = {Rev. Mod. Phys.},
	volume = {84},
	issue = {2},
	pages = {621--669},
	numpages = {0},
	year = {2012},
	publisher = {American Physical Society},
	doi = {10.1103/RevModPhys.84.621},
	url = {https://link.aps.org/doi/10.1103/RevModPhys.84.621}
}

@article{hotta2009quantum,
	title={Quantum energy teleportation in spin chain systems},
	author={Hotta, Masahiro},
	journal={Journal of the Physical Society of Japan},
	volume={78},
	number={3},
	pages={034001},
	year={2009},
	publisher={The Physical Society of Japan},
	doi = {10.1143/JPSJ.78.034001}
}

@book{NielsenChuang,
	title={Quantum Computation and Quantum Information},
	author={Nielsen, Michael A. and Chuang, Isaac L.},
	year={2010},
	publisher={Cambridge University Press},
	edition={10th Anniversary},
	isbn={9780521635035}
}

@article{RepeaterReview_Sangouard,
title = {Quantum repeaters based on atomic ensembles and linear optics},
author = {Sangouard, Nicolas and Simon, Christoph and de Riedmatten, Hugues and Gisin, Nicolas},
journal = {Rev. Mod. Phys.},
volume = {83},
issue = {1},
pages = {33--80},
numpages = {0},
year = {2011},
month = {Mar},
publisher = {American Physical Society},
doi = {10.1103/RevModPhys.83.33},
url = {https://link.aps.org/doi/10.1103/RevModPhys.83.33}
}

@Article{Ikeda2025,
	author    = {Ikeda, Kazuki},
	journal   = {Canadian Journal of Physics},
	title     = {Quantum energy in quantum computers: insights from quantum energy teleportation},
	year      = {2025},
	number    = {10},
	pages     = {939--948},
	volume    = {103},
	doi       = {10.1139/cjp-2025-0120},
	publisher = {Canadian Science Publishing},
}

@Article{Hotta2014,
	author    = {Hotta, Masahiro and Matsumoto, Jiro and Yusa, Go},
	journal   = {Physical Review A},
	title     = {Quantum energy teleportation without a limit of distance},
	year      = {2014},
	number    = {1},
	pages     = {012311},
	volume    = {89},
	doi       = {10.1103/PhysRevA.89.012311},
}

@Article{Xie2025,
	author    = {Xie, Songbo and Sajjan, Manas and Kais, Sabre},
	journal   = {Entropy},
	title     = {Strong {Local} {Passivity} in {Unconventional} {Scenarios}: {A} {New} {Protocol} for {Amplified} {Quantum} {Energy} {Teleportation}},
	year      = {2025},
	month     = nov,
	number    = {11},
	volume    = {27},
	doi       = {10.3390/e27111147},
}

@Article{Matsueda2025,
	author    = {Matsueda, Hiroaki and Masaki, Yusuke and Itoh, Kanji and Ono, Atsushi and Nasu, Joji},
	journal   = {Physical Review Research},
	title     = {Nonlocal correlations in quantum energy teleportation: {Perspectives} from their {Majorana} representations and information thermodynamics},
	year      = {2025},
	month     = aug,
	number    = {3},
	pages     = {033137},
	volume    = {7},
	doi       = {10.1103/1yr5-sn98},
}

@Article{Trevison2015,
	author    = {Trevison, Jose and Hotta, Masahiro},
	journal   = {Journal of Physics A},
	title     = {Quantum energy teleportation across a three-spin {Ising} chain in a {Gibbs} state},
	year      = {2015},
	number    = {17},
	pages     = {175302},
	volume    = {48},
	doi       = {10.1088/1751-8113/48/17/175302},
	publisher = {IOP Publishing},
}

@Article{SanchezCordova2024,
	author    = {Sánchez-Córdova, Mar and Berra-Montiel, Jasel},
	journal   = {The European Physical Journal Plus},
	title     = {Quantum energy teleportation in phase space quantum mechanics},
	year      = {2024},
	month     = dec,
	number    = {12},
	pages     = {1106},
	volume    = {139},
	doi       = {10.1140/epjp/s13360-024-05897-3},
}

@Article{Wu2024,
	author    = {Wu, Feng-Lin and Fan, Hao and Wang, Lu and Liu, Shu-Qian and Liu, Si-Yuan and Yang, Wen-Li},
	journal   = {Physical Review A},
	title     = {Strong local passive state in the minimal quantum energy teleportation model},
	year      = {2024},
	month     = jun,
	number    = {6},
	pages     = {062208},
	volume    = {109},
	doi       = {10.1103/PhysRevA.109.062208},
}

@Article{Hotta2010a,
	author    = {Hotta, Masahiro},
	journal   = {Journal of Physics A},
	title     = {Quantum energy teleportation with an electromagnetic field: discrete versus continuous variables},
	year      = {2010},
	month     = feb,
	number    = {10},
	pages     = {105305},
	volume    = {43},
	doi       = {10.1088/1751-8113/43/10/105305},
}

@Article{Hotta2010,
	author    = {Hotta, Masahiro},
	journal   = {Physics Letters A},
	title     = {Energy entanglement relation for quantum energy teleportation},
	year      = {2010},
	month     = jul,
	number    = {34},
	pages     = {3416--3421},
	volume    = {374},
	doi       = {10.1016/j.physleta.2010.06.058},
	publisher = {Elsevier BV},
}

@Article{Ikeda2024,
	author    = {Ikeda, Kazuki},
	journal   = {IET Quantum Communication},
	title     = {Long-range quantum energy teleportation and distribution on a hyperbolic quantum network},
	year      = {2024},
	number    = {4},
	pages     = {543--550},
	volume    = {5},
	doi       = {10.1049/qtc2.12090},
}

@Article{Yusa2011,
	author    = {Yusa, Go and Izumida, Wataru and Hotta, Masahiro},
	journal   = {Physical Review A},
	title     = {Quantum energy teleportation in a quantum {Hall} system},
	year      = {2011},
	month     = sep,
	number    = {3},
	pages     = {032336},
	volume    = {84},
	doi       = {10.1103/PhysRevA.84.032336},
}

@Article{Frey2013,
	author    = {Frey, Michael R. and Gerlach, Karl and Hotta, Masahiro},
	journal   = {Journal of Physics A},
	title     = {Quantum energy teleportation between spin particles in a {Gibbs} state},
	year      = {2013},
	month     = oct,
	number    = {45},
	pages     = {455304},
	volume    = {46},
	doi       = {10.1088/1751-8113/46/45/455304},
	publisher = {IOP Publishing},
}

@Article{Ikeda2023,
	author    = {Ikeda, Kazuki},
	journal   = {Physical Review D},
	title     = {Criticality of quantum energy teleportation at phase transition points in quantum field theory},
	year      = {2023},
	month     = apr,
	number    = {7},
	pages     = {L071502},
	volume    = {107},
	doi       = {10.1103/PhysRevD.107.L071502},
}

@Article{Fan2024,
	author    = {Fan, Hao and Wu, Feng-Lin and Wang, Lu and Liu, Shu-Qian and Liu, Si-Yuan},
	journal   = {Physical Review A},
	title     = {Strong quantum energy teleportation},
	year      = {2024},
	month     = nov,
	number    = {5},
	pages     = {052424},
	volume    = {110},
	doi       = {10.1103/PhysRevA.110.052424},
	publisher = {American Physical Society},
}

@Article{Ikeda2024a,
	author    = {Ikeda, Kazuki and Lowe, Adam},
	journal   = {Physical Review D},
	title     = {Robustness of quantum correlation in quantum energy teleportation},
	year      = {2024},
	month     = nov,
	number    = {9},
	pages     = {096010},
	volume    = {110},
	doi       = {10.1103/PhysRevD.110.096010},
	publisher = {American Physical Society},
}

@Article{Fan2024a,
	author    = {Fan, Hao and Wu, Feng-Lin and Wang, Lu and Liu, Shu-Qian and Liu, Si-Yuan},
	journal   = {Quantum Information Processing},
	title     = {The role of quantum resources in quantum energy teleportation},
	year      = {2024},
	month     = nov,
	number    = {11},
	pages     = {1},
	volume    = {23},
	doi       = {10.1007/s11128-024-04579-4},
}

@Article{Ikeda2023a,
	author    = {Ikeda, Kazuki},
	journal   = {Physical Review Applied},
	title     = {Demonstration of {Quantum} {Energy} {Teleportation} on {Superconducting} {Quantum} {Hardware}},
	year      = {2023},
	month     = aug,
	number    = {2},
	pages     = {024051},
	volume    = {20},
	doi       = {10.1103/PhysRevApplied.20.024051},
	publisher = {American Physical Society},
}

@Article{Wang2024,
	author    = {Wang, Jinzhao and Yao, Shunyu},
	journal   = {Quantum},
	title     = {Quantum {Energy} {Teleportation} versus {Information} {Teleportation}},
	year      = {2024},
	month     = dec,
	pages     = {1564},
	volume    = {8},
	doi       = {10.22331/q-2024-12-12-1564},
	publisher = {Verein zur Förderung des Open Access Publizierens in den Quantenwissenschaften},
}

@Article{Fiorini2025,
	author    = {Fiorini, Francesco and Garroppo, Rosario G. and Pagano, Michele},
	journal   = {IEEE Communications Magazine},
	title     = {Quantum {Repeaters} to {Extend} the {Communication} {Range}: {Present} and {Future} {Perspectives}},
	year      = {2025},
	pages     = {1--7},
	doi       = {10.1109/MCOM.001.2400724},
}

@Article{DiCandia2015,
	author    = {Di Candia, R. and Fedorov, K. G. and Zhong, L. and Felicetti, S. and Menzel, E. P. and Sanz, M. and Deppe, F. and Marx, A. and Gross, R. and Solano, E.},
	journal   = {EPJ Quantum Technology},
	title     = {Quantum teleportation of propagating quantum microwaves},
	year      = {2015},
	month     = dec,
	number    = {1},
	pages     = {25},
	volume    = {2},
	doi       = {10.1140/epjqt/s40507-015-0038-9},
}

@Article{Solmeyer2016,
	author    = {Solmeyer, Neal and Li, Xiao and Quraishi, Qudsia},
	journal   = {Physical Review A},
	title     = {High teleportation rates using cold-atom-ensemble-based quantum repeaters with {Rydberg} blockade},
	year      = {2016},
	month     = apr,
	number    = {4},
	pages     = {042301},
	volume    = {93},
	doi       = {10.1103/PhysRevA.93.042301},
	publisher = {American Physical Society},
}

@Article{Namiki2016,
	author    = {Namiki, Ryo and Jiang, Liang and Kim, Jungsang and L{\"u}tkenhaus, Norbert},
	journal   = {Physical Review A},
	title     = {Role of syndrome information on a one-way quantum repeater using teleportation-based error correction},
	year      = {2016},
	month     = nov,
	number    = {5},
	pages     = {052304},
	volume    = {94},
	doi       = {10.1103/PhysRevA.94.052304},
	publisher = {American Physical Society},
}

@Article{Goncharov2023,
	author    = {Goncharov, R. and Kiselev, Alexei D. and Moiseev, E. S. and Samsonov, E. and Moiseev, S. A. and Kiselev, F. and Egorov, V.},
	journal   = {Physical Review Applied},
	title     = {Quantum repeaters and teleportation via entangled phase-modulated multimode coherent states},
	year      = {2023},
	month     = oct,
	number    = {4},
	pages     = {044030},
	volume    = {20},
	doi       = {10.1103/PhysRevApplied.20.044030},
	publisher = {American Physical Society},
}

@Article{Ghosal2025,
	author    = {Ghosal, Arkaprabha and Ghai, Jatin and Saha, Tanmay and Ghosh, Sibasish and Alimuddin, Mir},
	journal   = {Physical Review Letters},
	title     = {Repeater-{Based} {Quantum} {Communication} {Protocol}: {Maximizing} {Teleportation} {Fidelity} with {Minimal} {Entanglement}},
	year      = {2025},
	month     = apr,
	number    = {16},
	pages     = {160803},
	volume    = {134},
	doi       = {10.1103/PhysRevLett.134.160803},
	publisher = {American Physical Society},
}

@Article{Wu2022,
	author    = {Wu, Bo-Han and Zhang, Zheshen and Zhuang, Quntao},
	journal   = {Quantum Science and Technology},
	title     = {Continuous-variable quantum repeaters based on bosonic error-correction and teleportation: architecture and applications},
	year      = {2022},
	month     = mar,
	number    = {2},
	pages     = {025018},
	volume    = {7},
	doi       = {10.1088/2058-9565/ac4f6b},
	publisher = {IOP Publishing},
}

@Article{Mylavarapu2025,
	author    = {Mylavarapu, Ganesh and Ghosh, Subrata and Hens, Chittaranjan and Chakrabarty, Indranil and Mitra, Subhadip},
	journal   = {Physical Review A},
	title     = {Teleportation fidelity of quantum repeater networks},
	year      = {2025},
	month     = sep,
	number    = {3},
	pages     = {032618},
	volume    = {112},
	doi       = {10.1103/72jp-37k6},
	publisher = {American Physical Society},
}

@Article{Dias2017,
	author    = {Dias, Josephine and Ralph, T. C.},
	journal   = {Physical Review A},
	title     = {Quantum repeaters using continuous-variable teleportation},
	year      = {2017},
	month     = feb,
	number    = {2},
	pages     = {022312},
	volume    = {95},
	doi       = {10.1103/PhysRevA.95.022312},
	publisher = {American Physical Society},
}
	
\end{document}